\documentstyle[psfig,12pt,a4wide]{article}

\newcommand{\news}{\setcounter{equation}{0}}

\def\eqn{\begin{equation}}
\def\eeqn{\end{equation}}
\def\arr{\begin{array}}
\def\earr{\end{array}}
\def\eqna{\begin{eqnarray}}
\def\eeqna{\end{eqnarray}}
\def\a{\alpha}
\def\b{\beta}
\def\d{\delta}
\def\D{\Delta}

\def\d{\delta}
\def\w{\wedge}

\def\r{\rho}
\def\O{\Omega}

\def\e{\epsilon}
\def\th{\theta}
\def\m{\mu}
\def\n{\nu}
\def\la{\lambda}
\def\La{\Lambda}
\def\z{\zeta}
\def\t{\tau}
\def\p{\partial}

\font\mybb=msbm10 at 12pt
\def\bb#1{\hbox{\mybb#1}}
\def\bZ {\bb{Z}}

\def\bE {\bb{E}}

\def\bM {\bb{M}}

\def\spa#1{\phantom{\fbox{\rule[-#1cm]{0cm}{0cm}}}}

\begin{document}

\vspace*{-.6in} \thispagestyle{empty}
\begin{flushright}
LPTENS--01--16\\
DAMTP--2001--32\\
\end{flushright}
\vspace{.2in} {\Large
\begin{center}
{\bf Flux--branes and the Dielectric Effect\\
in String Theory}
\end{center}}
\vspace{.2in}
\begin{center}
Miguel S. Costa$^{\dagger}$\footnote{miguel@lpt.ens.fr},\ \ Carlos
A.R. Herdeiro$^{\ddagger}$\footnote{car26@damtp.cam.ac.uk}\ \
and\ \ Lorenzo Cornalba$^{\dagger}$\footnote{cornalba@lpt.ens.fr}
\\
\vspace{.2in} $^{\dagger}$\emph{Laboratoire de Physique
Th\'eorique de
l'\'Ecole Normale Sup\'erieure\\
24 rue Lhomond, F--75231 Paris Cedex 05, France}\\
\vspace{.1in}
$^{\ddagger}$\emph{D.A.M.T.P. -- University of Cambridge\\
Center for Mathematical Sciences\\ Wilberforce Road, Cambridge CB3
0WA, UK}
\end{center}

\vspace{.3in}

\begin{abstract}
We consider the generalization to String and M-theory of the
Melvin solution. These are flux $p$--branes which have
$(p+1)$--dimensional Poincar\'e invariance and are associated to
an electric $(p+1)$--form field strength along their worldvolume.
When a stack of D$p$--branes is placed along the worldvolume of a
flux $(p+3)$--brane it will expand to a spherical D$(p+2)$--brane due
to the dielectric effect. This provides a new setup to consider the
gauge theory/gravity duality. Compactifying M--theory on a circle we
find the exact gravity solution of the type IIA theory 
describing the dielectric expansion of $N$ D4--branes into a
spherical bound state of D4--D6--branes, due to the presence of a
flux $7$--brane. In the decoupling limit, the deformation of the
dual field theory associated with the presence of the flux brane
is irrelevant in the UV. We calculate the gravitational radius and
energy of the dielectric brane which give, respectively, a
prediction for the VEV of scalars and vacuum energy of the dual field
theory. Consideration of a spherical D6--brane probe with $n$ units of 
D4--brane charge in the dielectric brane geometry suggests that the
dual theory  arises as the Scherk--Schwarz reduction of the
M5--branes $(2,0)$ conformal field theory. The probe potential has one
minimum placed at the locus of the bulk dielectric brane and another
associated to an inner dielectric brane shell.
\end{abstract}
\newpage

\section{Introduction}
\news
The realization that $p$--branes play a fundamental role in the
understanding of String and M-theory played a central role in the
developments of the past years. In particular, these $p$--branes
admit a gravitational description in the low energy supergravity
limit of String and M-theory \cite{HoroStro}. They can carry
either the magnetic or electric charges associated with the form
gauge potentials of these theories. With the discovery of
D--branes in perturbative string theory \cite{Polc:95} it was
possible to use string theory to study black holes
\cite{StroVafa}, which later led to the famous AdS/CFT duality
\cite{Mald:97,GKP,Witten}.

Another geometry that appears in the Einstein-Maxwell theory is
the so called Melvin Universe \cite{Melv:64} (see also $[8-11]$).
This geometry represents a magnetic flux tube in four dimensions.
In the cases that the gauge field arises from a Ka\l
u\.{z}a--Klein compactification the higher dimensional space for
the magnetic Melvin solution is flat with some non-trivial
identifications \cite{Dowker:93,Dowker:95,Dowker:96}. This fact led to a
construction of the type IIA flux--brane from the compactification
of M-theory on a circle \cite{RussoTsey,CostaGutp:01} (see also $[17-21]$
for further work on the Melvin solution in String theory). This
geometry has 8-dimensional Poincar\'e invariance and therefore is
a flux 7--brane (the Melvin fluxtube is a flux string in $4D$). It
has an additional $SO(2)$ invariance associated to the spherical
symmetry in the transverse plane to the brane. The natural
question that arises is to generalize this solution to the case of
forms with different rank in String and M-theory. This is one of
the purposes of this paper.

In analogy with the type IIA flux 7--brane, the flux $p$--branes
in $D$--dimensional space--time are associated to a flux of a
$(D-p-1)$--form field strength along the transverse space to the
brane. If one considers electric variables then they are
associated to an electric $(p+1)$--form field strength along the
brane worldvolume. This raises the issue of stability of the flux
$p$--branes. In fact, one expects to have Schwinger production of
spherical $(p-1)$--branes as shown in \cite{Dowker:96}. As such,
to find a string theory dual of the flux $p$--branes in the same
spirit of the AdS/CFT duality becomes problematic. However, we can
try to stabilize  the flux $p$--branes. Consider a Ramond--Ramond
flux $(p+3)$--brane and place on its worldvolume a stack of
D$p$--branes. Due to the coupling of the D$p$--branes to the
electric $(p+4)$--form field strength the brane will expand to a
dielectric 2--sphere forming a D$(p+2)$--brane. In other words,
the dielectric effect is at work \cite{Myers:99}. We shall argue
that, in the decoupling limit, the presence of the D$p$--branes
stabilizes the flux $(p+3)$--brane.

To be more precise let us briefly describe the work of Myers on
dielectric branes \cite{Myers:99}. Following general principles
such as gauge invariance and T--duality invariance, he found new
couplings to the bulk fields in the non--abelian D$p$--brane
action both in the Born--Infeld and in the Chern--Simons pieces of
the action. These new couplings would not be taken into account by
just replacing abelian by non--abelian gauge fields in a
D$p$--brane worldvolume action and by taking a gauge trace. In
particular, this work led to a new proposal for the Chern--Simons
term of the D$p$--brane action bosonic sector with the form 
\eqn
S_{CS}=\mu_p\int {\rm Tr}\left(P\left[e^{i2\pi\alpha'{\bf
i_{\Phi}} {\bf i_{\Phi}}} \left(\sum {\cal A}_n e^{-{\cal
B}}\right)\right]e^{2\pi\alpha' F}\right). 
\label{Chern-Simons}
\eeqn 
$P[\dots]$ denotes the pull--back of the background
Kalb--Ramond 2--form ${\cal B}$, and Ramond--Ramond $n$--form
potentials ${\cal A}_n$. The novelty of this proposal is the
exponential containing the operator ${\bf i_{\Phi}}$, denoting
interior multiplication by the transverse space vector ${\bf
\Phi}$ associated to displacements of the branes. Since this
reduces the degree of a differential form, it allows for
Wess--Zumino couplings of a D$p$--brane to Ramond--Ramond forms of
degree greater than $p+1$.

A physical effect arising from this new coupling can be seen by
considering a collection of $N$ D0--branes in a constant
background electric field described by a Ramond--Ramond 4--form
field strength. Then the scalar potential for this collection of
branes is 
\eqn 
V\left({\bf \Phi}\right)=-(2\pi\a')^2T_0
\left(\frac{1}{4}{\rm Tr}\left([ \Phi^i, \Phi^j]^2\right)
+\frac{i}{3}{\rm Tr}\left( \Phi^i \Phi^j \Phi^k\right) {\cal
F}_{tijk}\right)\ , 
\label{potential} 
\eeqn 
where $T_0$ is the
D0--brane mass and ${\cal F}_{tijk}=E\e_{ijk}$ along three
transverse directions. An analysis of the potential extrema shows
that the ground state is described by 
\eqn
\Phi^i=\frac{E}{4}\a^i\ , 
\label{SU(2)sphere} 
\eeqn 
where the
$\a^i$ belong to the $N\times N$ irreducible representation of the
$SU(2)$ algebra, with the value for the potential at large N reading 
\eqn 
V_N=-\frac{\pi^2\alpha'^{3/2}}{96g}E^4N^3\ .
\label{myerspot} 
\eeqn 
This non--commutative configuration
represents a single, somewhat granular, spherical D2--brane with
$N$ D0--branes bound to it. For large $N$ this sphere has a
`radius' 
\eqn 
r_s=\frac{\pi}{2}\alpha'|E|N\ . 
\label{myersrad}
\eeqn 
Of course, by T--duality a D$p$--brane immersed in a $(p+4)$
Ramond--Ramond electric field will expand to a D$(p+2)$--brane
with worldvolume geometry $\bM^{p+1}\times S^2$. The action of the
background electric field is to create a dipole moment (and higher
multipole moments) with respect to D$(p+2)$--brane charge. In
close analogy with classical electrodynamics Myers dubbed these
branes as Dielectric Branes. Now we see that the source for the
external electric field can be taken to be a flux brane.

In a beautiful paper \cite{PolcStra:00}, Polchinski and Strassler
realized that each of the vacua of the ${\cal N}=1^*$ Super
Yang--Mills theory, obtained by adding finite mass terms to the
${\cal N}=4$ theory, is dual to the gravity background associated
to a dielectric brane source (further work can be found in
$[24-27]$). In this general class of theories one can break all
the supersymmetry, for example, by adding a mass term to the
gluino. However, in constructing the gravitational dual, they
considered a spherical distribution of D3--branes where they
placed a D5--brane source, and then solved the gravity equations
to first order in the mass perturbation. A natural question that
arises is to fully describe the gravitational background for the
dielectric branes. We shall address this problem in this paper. By
placing a D$p$--brane in a Ramond--Ramond flux $(p+3)$--brane one
expects to generate a dielectric brane. The presence of the
D$p$--brane charge will stabilize the system. In fact, after
taking the decoupling limit this configuration is expected to be
stable since the dual field theory is on the general class of the
theories analyzed by Polchinski and Strassler. In particular, we
expect tachyons to be absent.

To construct the gravitational background for a dielectric brane,
the key idea is to consider the Ramond--Ramond magnetic flux
7-brane that arises from the reduction of M-theory flat
space--time with some non--trivial identifications
\cite{RussoTsey,CostaGutp:01}. We know that the double dimensional 
reduction of the M--theory 5--brane gives the type IIA D4--brane
\cite{Town:95}. Hence, it is natural to suspect that implementing
the Ka\l u\.{z}a--Klein Melvin twisted reduction to the M5--brane
will give a D4--brane in a magnetic flux 7-brane, or equivalently,
a D4--brane placed in a electric field. Then the coupling of the
D4--brane to the dual RR 7--form potential will give rise to the
Myers effect. Thus, after the reduction we expect to describe
within gravity a D4--brane expanded into a 2--sphere, i.e. a
D6--brane.

This paper is organized as follows. In section 2 we make the
ansatz for the flux $p$--branes. This geometry has the desired
$(p+1)$--dimensional Poincar\'e invariance together with
$SO(D-p-1)$ spherical symmetry. The associated $(p+1)$--form field
strength has a electric component along the brane worldvolume. The
corresponding system of differential equations does not decouple
in terms of non-interacting Liouville systems as it is usually the
case for black holes. We investigate the asymptotics of the flux
branes geometry.

In section 3 we construct the gravity solution for a D4-brane
expanded into a 2--sphere due to the dielectric effect. Firstly we
consider the twisted dimensional reduction of the non--extremal
M5--brane with a double analytic continuation. The
ten--dimensional configuration is interpreted as a D4--brane
immersed in a flux 7--brane with maximum magnetic field parameter.
Although this value of the magnetic field is unphysically large, we
chose, for the sake of clarity, to consider this case first since the
corresponding geometry has all the correct features to be interpreted
as a dielectric brane. Then we consider the general case of arbitrary
magnetic field which amounts to compactify a rotating M5--brane
after a similar double analytic continuation.

Section 4 is devoted to the study of the dielectric brane
geometry in the decoupling limit. The resulting geometry has the
same asymptotics as the D4-brane geometry in the decoupling limit,
which shows that the deformation of the dual theory associated to
the coupling of the D4--brane worldvolume theory to the flux
7--brane is irrelevant in the UV. Using the gravitational
description we calculate the scalars VEV and vacuum energy of the
field theory where the deformation becomes relevant. 

In section 5 we probe the dielectric brane geometry using a
spherical D6--brane probe with $n$ units of D4--brane charge. First we
consider the probe in the flux 7--brane background and obtain similar
results to those of Myers \cite{Myers:99}, but now taking into account
the backreaction of the background electric field on the geometry. This
simple case is very useful to understand the stability of the
dielectric brane geometry before taking the decoupling limit. We
proceed with the study of the probe in the dielectric brane geometry
in the decoupling limit. It is seen that either far away or inside the
dielectric brane the probe potential has the expected form to be
associated with the Scherk--Schwarz \cite{Scherk:1979} reduction of
the M5--branes $(2,0)$ low energy conformal field theory. This
potential has a minimum at the locus of the bulk dielectric brane and
another minimum in its interior that is associated to an inner
dielectric brane shell. 

We give our conclusions in section 6.

\section{Flux--branes}
\news

We shall consider the following general action in $D$--dimensional
space--time 
\eqn 
S=\frac{1}{2\kappa^2}\int d^Dx\sqrt{-g}
\left[R-\frac{1}{2}(\partial\phi)^2 -\frac{1}{2d!}e^{a\phi}{\cal
F}^2\right]\ , 
\label{action} 
\eeqn 
where $\kappa$ is the
gravitational coupling and ${\cal F}$ is a generic $d$--form field
strength. The above action can be regarded as a consistent
truncation of either String or M-theory low energy actions, where
${\cal F}$ represents any of the field strengths or
electromagnetic dual in these theories. Then the equations of
motion read 
\eqn 
\Box \phi = \frac{a}{2d!}e^{a\phi}{\cal F}^2\ ,\
\ \ \ \ \ d\left(e^{a\phi}\star {\cal F}\right)=0\ ,\ \ \ \ \ \
R_{ab}=\t_{ab}\ , 
\label{eqns} 
\eeqn 
where the tensor $\t=\t^{\phi}+\t^{{\cal F}}$ takes the form 
\eqn 
\arr{c}
\displaystyle{\t_{ab}^{\phi}=\frac{1}{2}\,\p_a\phi\,\p_b\phi}\ ,
\spa{0.4}\\
\displaystyle{\t_{ab}^{{\cal F}}=\frac{e^{a\phi}}{2}
\left[\frac{1}{(d-1)!}\,{\cal F}_{ac_1\cdots c_{d-1}} {\cal
F}_b^{\ c_1\cdots c_{d-1}} -\frac{d-1}{(D-2)d!}\,g_{ab}\,{\cal
F}^2\right]}\ . 
\earr 
\label{tau} 
\eeqn

In order to describe a flux--brane with $ISO(1,d-1)\times SO(D-d)$
invariance, we shall make the following ansatz 
\eqn 
\arr{c}
\displaystyle{ds^2=e^{2A(r)}ds^2\left(\bM^d\right)+e^{2C(r)}dr^2
+e^{2B(r)}E^{-2}d\O_{\tilde{d}}^{\ 2}}\ ,
\spa{0.4}\\
\displaystyle{{\cal F}=E\,\e\left(\bM^d\right)\ ,\ \ \ \ \
\phi=\phi(r)}\ , 
\earr 
\label{ansatz} 
\eeqn 
where
$\e\left(\bM^d\right)$ is the volume form of $d$--dimensional
Minkowski space $\bM^d$ and $d\O_{\tilde{d}}$ is the metric
element on the unit $\tilde{d}=(D-d-1)$--sphere. We have
conveniently multiplied the line element of this sphere by
$E^{-1}$ for dimensional reasons. Let us note that the form ${\cal
F}$ is closed since $E$ is constant and also automatically solves
the corresponding equation of motion. Notice that we refer to this
electric field as constant due to the independence on the radial
coordinate (however it is not covariantly constant).
Alternatively, one could consider the electromagnetic dual of
${\cal F}$ which reads 
\eqn 
\tilde{{\cal F}}=E^{1-\tilde{d}}\,e^{-dA+C+\tilde{d}B}\,dr
\w\e\left(S^{\tilde{d}}\right)\ , 
\label{dual} 
\eeqn 
where $\tilde{{\cal F}}$ is a $(D-d)$--form
and $\e\left(S^{\tilde{d}}\right)$ is the volume form on the unit
$\tilde{d}$--sphere.

The Ricci tensor for the above geometry is 
\eqn 
\arr{c}
\displaystyle{R_{\m\n}=
-e^{2A-2C}\left[A''+d(A')^2-A'C'
+\tilde{d}A'B'\right]\eta_{\m\n}}\ ,
\spa{0.4}\\
\displaystyle{R_{ij}=
-e^{2B-2C}\left[B''+\tilde{d}(B')^2-B'C'+dA'B'\right]h_{ij}
+E^2\left(\tilde{d}-1\right)h_{ij}}\ ,
\spa{0.4}\\
\displaystyle{R_{rr}= \tilde{d}\left[B'C'-B''-(B')^2\right]
+d\left[A'C'-A''-(A')^2\right]}\ , 
\earr 
\label{Ricci} 
\eeqn 
where $'$ denotes differentiation with respect to the radial
coordinate $r$. We are using coordinates $x^{\m}$ along the
worldvolume directions of the brane $\bM^d$ and $\th^i$ on the
$\tilde{d}$--dimensional unit sphere $S^{\tilde{d}}$. The only
non--vanishing component of the tensor $\t^{\phi}$ is 
\eqn
\t^{\phi}_{rr}=\frac{1}{2}(\phi')^2\ , 
\label{tauphi} 
\eeqn 
and those of $\t^{{\cal F}}$ are 
\eqn 
\arr{c} 
\displaystyle{\t^{\cal
F}_{\m\n}= -\frac{E^2}{2}\,e^{a\phi}
\left(\frac{\tilde{d}}{D-2}\right)\,e^{-2(d-1)A}\,\eta_{\m\n}}\ ,
\spa{0.6}\\
\displaystyle{\t^{\cal F}_{ij}= \frac{E^2}{2}\,e^{a\phi}
\left(\frac{d-1}{D-2}\right)\,e^{2B-2dA}\,h_{ij}}\ ,
\spa{0.6}\\
\displaystyle{\t^{\cal F}_{rr}= \frac{E^2}{2}\,e^{a\phi}
\left(\frac{d-1}{D-2}\right)\,e^{2C-2dA}}\ . 
\earr 
\label{tauF}
\eeqn

In the ansatz (\ref{ansatz}) there is a freedom of reparametrization
of the radial coordinate $r$. We conveniently choose the gauge
\eqn 
dA+\tilde{d}B=C\ , 
\label{gauge} 
\eeqn 
in which the equations of motion simplify to 
\eqn 
\arr{c}
\displaystyle{\phi''=-a\frac{E^2}{2}\,e^{a\phi+2\tilde{d}B}}\ ,
\spa{0.5}\\
\displaystyle{A''=\frac{E^2}{2}\left(\frac{\tilde{d}}{D-2}\right)\,
e^{a\phi+2\tilde{d}B}}\ ,
\spa{0.6}\\
\displaystyle{B''=-\frac{E^2}{2}\left(\frac{d-1}{D-2}\right)\,
e^{a\phi+2\tilde{d}B}+
E^2\left(\tilde{d}-1\right)\,e^{2dA+2(\tilde{d}-1)B}}\ , 
\earr
\label{eqns2} 
\eeqn 
together with a {\em zero--energy} constraint
from the $rr$ component of Einstein equations. This constrained
system of differential equations can be derived from the
Lagrangian ${\cal L}=T-V$ where 
\eqn
T=-\frac{1}{2}(\phi')^2+d(d-1)(A')^2+\tilde{d}(\tilde{d}-1)(B')^2
+2d\tilde{d}A'B'\ , \label{T} \eeqn and \eqn
V=-E^2\tilde{d}(\tilde{d}-1)e^{2dA+2(\tilde{d}-1)B}-
\frac{E^2}{2}e^{2\tilde{d}B+a\phi}\ , 
\label{V} 
\eeqn 
with the {\em zero--energy} constraint $T+V=0$. In the case of black  
holes one can usually decouple this system in terms of non-interacting
Liouville systems related through the zero energy condition (see
for example \cite{Callan}). However, in this case it is not
possible to decouple the system, which makes the solution of the
problem much harder. For this reason we were not able to find a
general analytic solution but will investigate the asymptotics of
the flux branes.

To analyze the differential equation (\ref{eqns2}) we define 
\eqn
\arr{c} 
\displaystyle{f=a\,\phi+2\tilde{d}\,B}\ ,
\spa{0.2}\\
\displaystyle{g=2d\,A+2\left(\tilde{d}-1\right)B}\ ,
\spa{0.3}\\
\displaystyle{h=\left(\frac{\tilde{d}}{D-2}\right)\phi+a\,A}\ .
\earr 
\label{functions} 
\eeqn 
These functions satisfy the following system of differential equations 
\eqn 
\arr{c} 
h''=0\ ,
\spa{0.2}\\
f''=c_1\,e^f+c_2\,e^g\ ,
\spa{0.2}\\
g''=c_3\,e^f+c_4\,e^g\ , 
\earr 
\label{fgeqns} 
\eeqn 
where the constant coefficients $c_i$ have the form 
\eqn 
\arr{c}
\displaystyle{c_1=
-\frac{E^2}{2}\left[a^2+2\frac{\tilde{d}(d-1)}{D-2}\right] \equiv
-\la^2}\ ,
\spa{0.5}\\
\displaystyle{c_2=2\tilde{d}\left(\tilde{d}-1\right)E^2\ ,\ \ \ \
c_3=E^2\ ,\ \ \ \ c_4=2\left(\tilde{d}-1\right)^2E^2}\ . 
\earr
\label{coefficients} 
\eeqn 
We shall see that, in the cases of
interest in String and M--theory, the constant $c_1$ simplifies to
$c_1=-2E^2$.

\subsection{Dilatonic Melvin}

For the usual dilatonic Melvin solution \cite{GibbMaeda} one has
$\tilde{d}=1$ and the transverse sphere is a circle which is flat.
As a consequence $c_2=c_4=0$ and the equations (\ref{fgeqns})
simplify considerably. In this case $c_3f''=c_1g''$, and the
general solution is given by 
\eqn 
e^f=\left(\frac{\a_1}
{\cosh{\left[\frac{\a_1}{\sqrt{2}}(\la r+\a_2)\right]}}\right)^2\ , 
\ \ \ \ \ 
g=\frac{c_3}{c_1}\,f+\a_3+\a_4\,Er\ ,
\ \ \ \ \
h=\a_5+\a_6\,Er\ , 
\label{melvin} 
\eeqn 
where the $\a_i$'s are
dimensionless constants of integration. Notice that not all these
constants are independent because of the {\em zero--energy}
constraint.

The case we consider here is that of the RR flux 7-brane of type
IIA strings studied in \cite{CostaGutp:01}. In this case we have
$a=-3/2$ and $d=8$. This solution corresponds to choosing the
constants of integration in (\ref{melvin}) such that 
\eqn
e^f=\frac{1}{\cosh^2{(Er)}}\ ,\ \ \ \ \ e^g=1+e^{2Er}\ ,
\ \ \ \ \ h=0\ . 
\label{solf7} 
\eeqn 
Then the coordinate transformation
$Er=\ln{(E\rho/2)}$ will bring the metric, field strength and
dilaton field to the form 
\eqn 
\arr{c} 
\displaystyle{ ds^2
=\La^{1/8}\left[ds^2\left(\bM^8\right)+d\rho^2\right]
+\La^{-7/8}\rho^2\,d\varphi^2\ ,}
\spa{0.4}\\
\displaystyle{{\cal F}=E\,\e\left(\bM^8\right)\ ,\ \ \ \ \
e^{4\phi/3}=\La \equiv 1+(E\rho)^2/4}\ . 
\earr 
\label{sol2f7}
\eeqn 
The space--time metric is for $\rho E\ll 1$ approximately
flat. We can regard this as the boundary conditions for the flux
7-brane which fixes the constants of integration in
(\ref{melvin}). This fact is also true for the other flux branes
and was explored in \cite{GutpStro}.

The general solution (\ref{melvin}) contains singular geometries
for which the Ricci scalar blows up. These are naked
singularities. Only for the flux 7--brane solution (\ref{sol2f7})
is the geometry non-singular. One should regard the flux--branes
as non-supersymmetric vacua where the energy density of the
electromagnetic field spreads to infinity. In this sense the flux
branes do not represent localized lumps of energy and are not
asymptotically Minkowskian as the usual branes. Far away the
energy associated with the constant electric field will dominate.

To analyse the geometry (\ref{sol2f7}) we should multiply the
metric by the conformal factor $e^{\phi/2}$ to change to the
string frame. It turns out that this geometry is quite different
than the usual 4D Melvin Universe \cite{Melv:64}. In the latter
the orbits of $\partial/\partial \phi$ have vanishing length at
large radial distance. In the former type IIA case the length of
the $\partial / \partial \phi$ orbits $l_{\phi}$ scales in terms
of proper radial distance $u$ as $l_{\phi} \sim  u^{1/3}$. This
means that as $u\rightarrow\infty$ space-time does not close.
Also, this means that while in the 4D Melvin we have a
quantization condition for the flux of $\star {\cal F}$ through
the transverse space \cite{Gibbons:2001}, this no longer happens
for the IIA flux 7--brane (\ref{sol2f7}).

\subsection{Flux--branes in type II Strings}

Now we turn to the flux branes of type II String theory. We shall
consider the case of RR flux branes. In this case the coupling
$a=(5-d)/2$, $\tilde{d}=9-d$ and the coefficient $c_1$ in
(\ref{coefficients}) simplifies considerably to $c_1=-2E^2$. The
cases of NSNS flux branes can be obtained by a S-duality
transformation on the IIB flux 2-- and 6--branes.

For the RR flux branes the metric functions and dilaton are
related to the functions $f$, $g$ and $h$ by the expression
(\ref{functions}) that reads 
\eqn 
\arr{c}
\displaystyle{f=\left(\frac{5-d}{2}\right)\phi+2(9-d)\,B}\ ,
\spa{0.5}\\
\displaystyle{g=2d\,A+2(8-d)\,B}\ ,
\spa{0.2}\\
\displaystyle{h=\left(\frac{9-d}{8}\right)\phi
+\left(\frac{5-d}{2}\right)A}\ . 
\earr 
\label{RRfunctions} 
\eeqn
The functions $f$ and $g$ satisfy the system of differential
equations 
\eqn 
\arr{c} 
f''=-2E^2\,e^f+2(9-d)(8-d)E^2\,e^g\ ,
\spa{0.2}\\
g''=E^2\,e^f+2(8-d)^2E^2\,e^g\ . 
\earr 
\label{RRfgeqns} 
\eeqn 
A particular solution to this system of equations can be found by
setting $e^g=\z e^f$ \cite{Saffin}. Then from $f''=g''$ we must have 
\eqn 
\z=\frac{3}{2(8-d)}\ . 
\label{la} 
\eeqn 
Solving for $f$ we find 
\eqn 
e^f=\frac{2}{(25-3d)}\frac{1}{(Er)^2}\ , 
\label{RRf}
\eeqn 
and we also set $h=0$. A rather tedious calculation gives
the functions $A$, $B$, and $C$ that appear in the Einstein metric
(\ref{ansatz}) 
\eqn 
\arr{l} 
\displaystyle{
2A=\frac{(9-d)^2}{8(25-3d)}\,\ln\z+\frac{9-d}{8(25-3d)}\,f}\ ,
\spa{0.5}\\
\displaystyle{
2B=\frac{(d-5)^2}{8(25-3d)}\,\ln\z+\frac{25-d}{8(25-3d)}\,f}\ ,
\spa{0.5}\\
\displaystyle{2C=\frac{(9-d)(25-d)}{8(25-3d)}\,\ln\z+
\frac{25(9-d)}{8(25-3d)}\,f}\ . 
\earr 
\label{RRABC} 
\eeqn 
This solution satisfies the {\em zero--energy} constraint defined
above. We can change to the
string frame, and write the metric in coordinates such that the
radial coordinate is the proper radial distance. The final result
for the metric and dilaton field is 
\eqn 
\arr{c}
\displaystyle{ds^2=\b^{1/5}(Eu)^{2/5} ds^2\left(\bM^d\right)+du^2+
\frac{\b}{\z}\,u^2\,d\O_{9-d}^{2}}\ ,
\spa{0.6}\\
\displaystyle{e^{\frac{10}{d-5}\phi}=\z(Eu)^2}\ . 
\earr
\label{RRfluxbrane}
\eeqn 
where 
\eqn 
\b=\frac{5^2}{2^3(25-3d)}\ .
\eeqn 
This metric describes a geometry with a naked singularity at
the origin. However, its asymptotics are those of the RR flux
$(d-1)$--brane \cite{Saffin,GutpStro}, for which we expect a
smooth geometry. Far away the energy density associated with the
electric field dominates and determines the
asymptotics\footnote{Notice that for $d<8$ the powers of $u$ in
the metric (\ref{RRfluxbrane}) are independent of $d$. On the
other hand for $d=8$, the Melvin case, the powers of $u$ are
different.}. A comparison with flat space--time is helpful; for
example, the Schwarzchild black hole with negative mass has a
naked singularity and  is unphysical, however, it converges to
flat space--time at infinity where the energy density vanishes.
Similarly the solution (\ref{RRfluxbrane}) has the correct
asymptotics for the flux $(d-1)$--brane. For $d<5$, i.e. for the
flux $p$-branes with $p<4$, the dilaton field converges to zero at
infinity and we are in the perturbative string theory regime. For
$d>5$, the string coupling diverges. The case $d=5$ gives the
non--dilatonic (and self--dual) flux $4$--brane of the type IIB
theory. This is analogous to what happens with the D3--brane.
Finally, notice that while for the flux 7--brane the flux of
$\star {\cal F}$ along the transverse space is convergent, for the
flux branes analyzed here it diverges.

\subsection{Flux--branes in M-theory}

In M-theory there are flux 3-- and 6--branes, both non--dilatonic.
We have $D=11$ and $a=0$. The functions $f$ and $g$ are related to
those appearing in the metric by 
\eqn 
\arr{c} 
f=2(10-d)B\ ,
\spa{0.2}\\
g=2d\,A+2(9-d)B\ . 
\earr 
\label{Mfunctions} 
\eeqn 
These functions satisfy the system of differential equations 
\eqn 
\arr{c}
\displaystyle{f''=-2E^2\,e^f+2(10-d)(9-d)E^2\,e^g}\ ,
\spa{0.2}\\
\displaystyle{g''=E^2\,e^f+2(9-d)^2E^2\,e^g}\ . 
\earr
\label{Mfgeqns} 
\eeqn 
To find the asymptotics of the M
flux--branes we set $e^g=\eta e^f$, which gives 
\eqn
\eta=\frac{3}{2(9-d)}\ . 
\label{eta} 
\eeqn

In the case of the flux 3--brane $(d=4)$ the function $f$ reads
\eqn 
e^f=\frac{1}{8(Er)^2}\ , 
\label{M3f} 
\eeqn 
and the metric functions $A$, $B$, and $C$ become 
\eqn
2A=\frac{1}{4}\,\ln\eta+\frac{1}{24}\,f\ ,\ \ \ \ \
2B=\frac{1}{6}\,f\ ,\ \ \ \ \ 2C=\ln\eta+\frac{7}{6}\,f\ .
\label{M3ABC} 
\eeqn 
This gives the following metric written in
terms of the proper radial distance coordinate $u$: 
\eqn
ds^2=\left(\frac{2}{9}\right)^{1/4}(Eu)^{1/2}
ds^2\left(\bM^4\right)+du^2+\frac{20}{27}\,u^2\,d\O_6^{\ 2}\ .
\label{M3} 
\eeqn

For the flux 6--brane $(d=7)$ we have 
\eqn 
e^f=\frac{2}{7(Er)^2}\ , 
\label{M6f} 
\eeqn 
and 
\eqn
2A=\frac{1}{7}\,\ln\eta+\frac{1}{21}\,f\ ,\ \ \ \ \
2B=\frac{1}{3}\,f\ ,\ \ \ \ \ 2C=\ln\eta+\frac{4}{3}\,f\ .
\label{M6ABC} 
\eeqn 
The corresponding metric element can be written in the form 
\eqn
ds^2=\left(\frac{7}{18}\right)^{1/7}(Eu)^{2/7}
ds^2\left(\bM^7\right)+du^2+\frac{14}{27}\,u^2\,d\O_3^{\ 2}\ .
\label{M6} 
\eeqn 
As for the flux $p$--branes of String theory with
$p<7$, the flux along the transverse space diverges for the M
flux--branes.

\subsection{Stability of the flux--branes}

The flux $p$--branes described above are not stable. They will
decay through the nucleation of spherical $(p-1)$--branes as
described generally in \cite{Dowker:96}. If we compactify $(p-1)$
directions of the flux $p$--branes then they will decay through
the usual Schwinger production of $(p-1)$ brane/anti--brane pairs.
Similarly to the RR flux $7$--brane case \cite{CostaGutp:01}, one
can consider a $(p-1)$--brane probe at the core of the flux
$p$--brane and calculate the action for the instanton associated
with this decay process. This gives the well known result for the
nucleation rate $\Gamma$ 
\eqn 
\Gamma\sim e^{-I}\ ,\ \ \ \ \
I=\pi\frac{M_{p-1}}{|E|}\ , 
\eeqn 
where $M_{p-1}$ is the mass of a
$(p-1)$--brane. It is expected that this calculation can be
reproduced using the Euclidean Quantum gravity approximation for
the nucleation of brane pairs. One would need to find the
instanton for the nucleation process with the same asymptotics of
the flux--branes described above. This calculation could confirm
the expected periodicity of the electric (or magnetic) field
parameter. As explained in \cite{CostaGutp:01}, the existence of a
maximum electric field for a generic $p$--form is expected on the
basis of String duality and of the analysis of the RR flux
$7$--brane case from a M-theory perspective. An interesting
physical interpretation for this maximum electric field was given
in \cite{GutpStro}: since the typical distance for nucleation is
of order $1/E$, for larger values of E the black hole horizons
will touch and the pair production will cease to exist.

A very interesting question is to consider the string theory duals
of the flux brane geometries in some decoupling limit. This can be
problematic because, as explained above, these geometries are not
stable, which makes the duality difficult to establish (see
\cite{GutpStro} for a discussion of this point). However, one can
try to stabilize the flux branes. As explained in the Introduction
this can be done by considering the dielectric effect in String
theory \cite{Myers:99}.

Consider the case of a RR flux $(p+3)$--brane and place a stack of
$N$ D$p$-branes along its worldvolume. Then the D$p$-branes will
couple to the electric RR $(p+4)$-form field strength expanding
into a D$(p+2)$-brane with geometry $\bM^{p+1}\times S^2$. Now,
the presence of the $N$ D$p$--branes changes the asymptotics of
the geometry. Far away, we have the geometry for the flux--brane
together with a charge due to the $N$ D$p$--branes. We would then
need to find an instanton with these asymptotics, representing the
instability of space--time. We know from the perturbative String
theory description of the dielectric effect that this system is
locally stable, and therefore such an instanton represents a
quantum tunneling effect. Moreover, we shall argue that, the
geometry in the decoupling limit has the same asymptotics as the
usual D--branes without external electric field. Therefore, in
this limit the instanton instability no longer exists and the
configuration is stable. Note that the case of $p=3$ is nothing
but a D3--brane expanding to a spherical D5--brane due to the
dielectric effect. These type of configurations have already made
their appearance in the gauge theory/gravity duality of the
theories of the type analyzed by Polchinski and Strassler
\cite{PolcStra:00}, which are stable.

In the following sections we shall treat the case of $p=4$, where one
can find the exact gravitational description by using the M-theory
reduction of the M5--brane to the type IIA theory and hence
confirming the aforementioned expectations.

\section{Dielectric branes}
\news

Because of the complexity of the gravitational background
presented below it is important to set our conventions for the
bosonic sector of the  eleven--dimensional supergravity action:
\eqn 
{\cal S}=\frac{1}{2\kappa_{11}^{\ \ 2}}\left\{\int d^{11} x
\sqrt{-g}\left[R-\frac{1}{2\cdot 4!}{\cal F}^2\right]
+\frac{1}{6}\int {\cal F}\w {\cal F}\w {\cal A}\right\}\ ,
\label{11Daction} 
\eeqn 
where $\kappa_{11}$ is the
eleven--dimensional gravitational coupling and ${\cal F}=d{\cal
A}$ with ${\cal A}$ a 3--form field potential. Reduction to the
type IIA theory is achieved through the ansatz 
\eqn 
ds_{11}^{\ \ 2}=
e^{-2\phi/3}ds_{10}^{\ \ 2}+ e^{4\phi/3}\left(dx^{11}
+{\cal A}_adx^a\right)^2\ ,
\ \ \ \ \ 
{\cal A}={\cal A}_3+{\cal B}\w dx^{11}\ . 
\label{kkansatz} 
\eeqn 
We shall present the
construction of the dielectric branes in two steps. First we
consider a D4--brane placed in a flux 7--brane with maximum
magnetic (or electric) field. This value of the magnetic field is
unphysically large but the solution is simpler and retains all the
correct features. Then we consider the case of arbitrary magnetic
field.

\subsection{Maximal magnetic field}

Consider the non--extremal M5--brane solution obtained by a double
analytic continuation from the usual solution. The metric and
3--form potential read 
\eqn 
\arr{c} 
\displaystyle{ds_{11}^{\ \ 2}=
H^{-1/3}\left[ds^2\left(\bM^{5}\right)+fd\t^2\right]+}
\spa{0.4}\\
\displaystyle{
+H^{2/3}\left[\frac{dr^2}{f}+r^2\left(d\th^2+\sin^2{\th}\,d\tilde{\varphi}^2
+\cos^2{\th}\,d\O_2^{\ 2}\right)\right]}\ ,
\spa{0.6}\\
{\cal A}=-r_{H}^{\ 3}\sinh{\a}\,\cosh{\a}\,\cos^3{\th}\,
d\tilde{\varphi}\w\e\left(S^2\right)\ , 
\earr 
\label{EuclideanM5}
\eeqn where $0\le\tilde{\varphi}<2\pi$, $0\le\th\le\pi/2$. The
unusual parameterization of the transverse 4--sphere is standard
for rotating black holes \cite{PerryMyers}. In this coordinate
system the transverse space naturally splits into the $\th=0$
three--plane orthogonal to the $\th=\pi/2$ two--plane. The
functions $f$ and $H$ have the form 
\eqn
H=1+\left(\frac{R}{r}\right)^3\equiv
1+\left(\frac{r_H}{r}\right)^3\sinh^2{\a}\ ,\ \ \ \ \
f=1-\left(\frac{r_H}{r}\right)^3\ . 
\label{Hf} 
\eeqn 
The M5--brane charge quantization gives the condition 
\eqn 
r_H^{\ 3}\sinh{\a}\,\cosh{\a}=\pi N l_P^{\ 3}\ , 
\label{Nquanta} 
\eeqn
where $l_P$ is the eleven--dimensional Planck length and $N$ the
number of M5--branes. In order to avoid a conical singularity, the
Euclidean time direction $\t$ has periodicity given by 
\eqn 
2\pi
R_{11}=\frac{4\pi}{3}\,r_H \cosh{\a}\ , \label{periodtau} 
\eeqn
and it is related to the ten--dimensional type IIA string coupling
and tension by $R_{11}=g\sqrt{\a'}$.

If we compactify along the killing vector ${\p }/{\p\t}$ there
will be a set of fixed points at $r=r_H$, spanning a 4--sphere in
the transverse space. The reduced space will be singular on such a
4--sphere. One can instead compactify along the killing vector
field 
\eqn 
q=\frac{\p\ }{\p\t}+B\frac{\p\ }{\p\tilde{\varphi}}\ ,\
\ \ \ B=\frac{1}{R_{11}}\ , 
\label{compactvect} 
\eeqn 
which corresponds to the maximum value for the magnetic field. Notice
that the parameter $B$ is related to the electric field $E$ in the
two previous sections by $E=2B$. The fixed points of this
isometry are at $r=r_H$, $\th=0$ corresponding to a 2--sphere on
the transverse space (see figure 1). At each point of this
2--sphere the action of the isometry is the same as for a Ka\l
u\.{z}a--Klein monopole. Hence, the reduced space will be singular
on a 2--sphere that we identify with the D4--branes expanded into
a D6--brane. Asymptotically space--time will look like the flux
7--brane with maximal magnetic field parameter $B=1/R_{11}$,
together with the D4--brane charge. We have chosen this particular
value for $B$ because, as will be seen below, any other choice
would lead to a conical singularity for the ten--dimensional
geometry. For this value of the magnetic field, perturbative
string theory will hold only for $r\ll R_{11}$, while the eleventh
direction remains unobservable for $r\gg R_{11}$. We shall
consider this unphysical case first because it is much simpler and
retains all the features of the general case. In the next
subsection we shall allow for general and physical values of the
magnetic field by considering the rotating M5--brane geometry.

\begin{figure}
\begin{picture}(0,0)(0,0)
\put(186,22){$S^2_{r_H}$} \put(200,54){$_{\theta=0}$}
\put(106,113){$\tilde{r},\tau$} \put(20,17){$S^4$}
\put(95,70){$\tau$} \put(80,90){$\tilde{r}$}
\put(232,-6){$S^2_{r_H\cos{\theta}}$} \put(234,100){$\theta$}
\put(361,123){$_{\theta=\pi/2}$} \put(391,35){$\tilde{\varphi}$}
\put(289,23){$_{\bullet}$} \put(222,54){$_{\bullet}$}
\put(59,56){$_{\bullet}$}
\end{picture}
\centering\psfig{figure=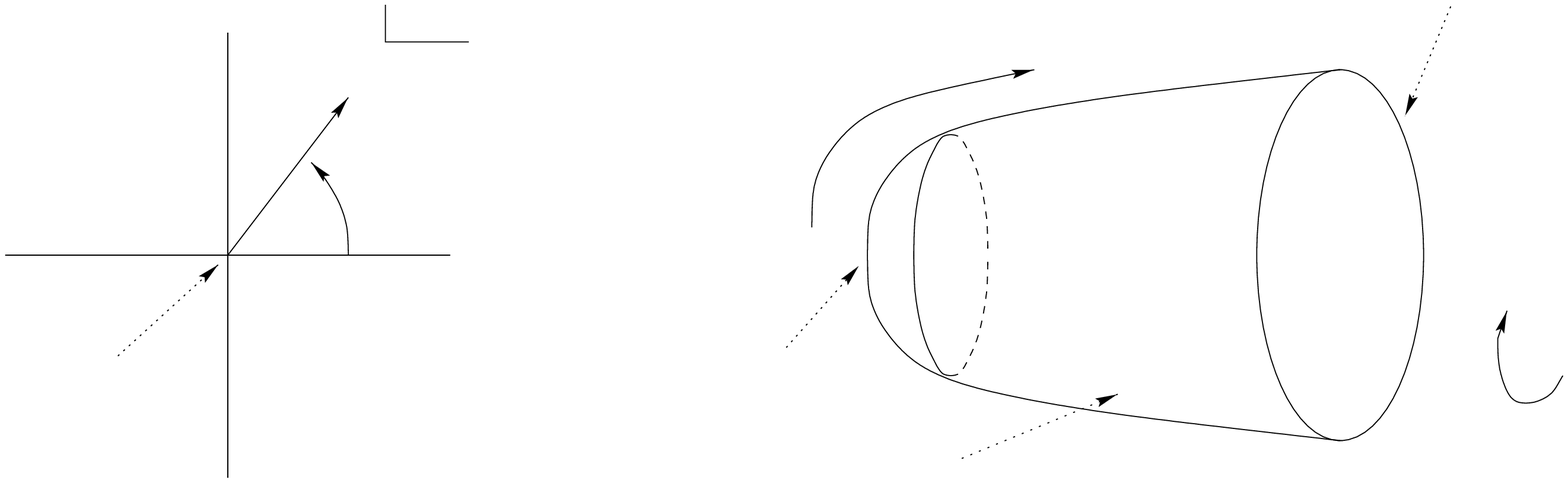,width=14cm} \caption{\small{Left:
The $(\tilde{r},\t)$ Euclidean plan with $\tilde{r}^2=r^2-r_H^{\
2}$. At $\tilde{r}=0$ there is a 4--sphere of radius $r_H$; Right:
The parameterization of the 4--sphere; the subscript on the $S^2$
denotes the radius.}} \label{fig1}
\end{figure}

To perform the reduction we change to the azimuthal angle
$\varphi=\tilde{\varphi}-B\t$, then $\t$ has period $2\pi R_{11}$
and $\varphi$ has the standard period of $2\pi$. A straightforward
calculation gives the following type IIA background fields: 
\eqn
\arr{c} 
\displaystyle{ds_{10}^{\ \ 2}=
\left(\frac{\Sigma}{H}\right)^{1/2} ds^2\left(\bM^{5}\right)
+\left(\Sigma H\right)^{1/2}\left[\frac{dr^2}{f}+
r^2\left(d\th^2+\cos^2{\th}\,d\O_2^{\ 2}\right)\right]+}
\spa{0.5}\\
\displaystyle{+ \left(\frac{H}{\Sigma}\right)^{1/2}
f\,r^2\sin^2{\th}\,d\varphi^2}\ ,
\spa{0.5}\\
\displaystyle{e^{2\phi}=\Sigma^{3/2}H^{-1/2}}\ ,\ \ \ \ \
\displaystyle{{\cal B}= -r_H^{\
3}B\sinh{\a}\,\cosh{\a}\,\cos^3{\th}\,\e\left(S^2\right)}\ ,
\spa{0.3}\\
\displaystyle{{\cal A}_1=B\Sigma^{-1}Hr^2\sin^2{\th}\,d\varphi}\ ,
\ \ \ \ \ \displaystyle{{\cal A}_3=-r_H^{\
3}\sinh{\a}\,\cosh{\a}\,\cos^3{\th}\,
d\varphi\w\e\left(S^2\right)}\ . 
\earr 
\label{gravdiele} 
\eeqn 
The function $\Sigma$ has the form 
\eqn 
\Sigma\equiv
f+H(Br\sin{\theta})^2\ , 
\label{Sigma} 
\eeqn 
and $B=1/R_{11}=1/(g\sqrt{\a'})$. Notice that $g$ is the asymptotic
value of the string coupling along the $\th=0$ three--plane.

\subsubsection{The ten--dimensional geometry}

The above solution describes a geometry with 5--dimensional
Poincar\'e invariance as it is appropriate to describe a
D4--brane. In fact, by integrating ${\cal F}_4$ over the
four--sphere we obtain the total D4--brane quanta of charge 
\eqn
N=\frac{1}{(2\pi)^3g\a'^{3/2}}\int {\cal F}_4= \frac{r_H^{\
3}\sinh{\a}\,\cosh{\a}}{\pi g \a'^{3/2}}\ . 
\label{N} 
\eeqn

To interpret this geometry consider first its asymptotics. For
$r\gg R,\,r_H$ we obtain the metric, dilaton field and 1--form
potential for the flux 7--brane solution\footnote{This solution
can be brought to the form of that in section two by the
coordinate transformation $v=r\cos{\th}$ and $\rho=r\sin{\th}$.}
\eqn 
\arr{c} 
\displaystyle{ds_{10}^{\ \ 2}=\La^{1/2}
\left\{ds^2\left(\bM^{5}\right)+dr^2+
r^2\left(d\th^2+\cos^2{\th}\,d\O_2^{\ 2}\right)
+\frac{r^2\sin^2{\th}}{\La}\,d\varphi^2\right\}}\ ,
\spa{0.5}\\
\displaystyle{e^{2\phi}=\La^{3/2}\ , \ \ \ \ {\cal
A}_1=r^2\sin^2{\th}\,B\La^{-1}d\varphi\ ,} 
\earr 
\label{Melvin}
\eeqn 
where $\La=1+(Br\sin{\th})^2$. A simple calculation shows
that the dual of the 2--form RR field strength is 
\eqn 
{\cal F}_8=2B\,\e\left(\bM^{5}\right)\w \e\left(\bE^{3}\right)\ ,
\label{dualF} 
\eeqn 
where $\e(\bE^{3})$ is the volume form on the
$\th=0$ three--plane. This supports the interpretation of the
solution as a D4--brane immersed in a constant RR 8--form electric
field, i.e. on a flux 7--brane.

Coming in from the large $r$ region one first finds a metric
singularity at $r=r_{H}$. But unlike the usual D4--brane solution,
the immersion in the magnetic flux brane makes the geometry
smoother on this hypersurface. Evaluating the Ricci scalar of
(\ref{gravdiele}) on the horizon $r=r_H$ yields 
\eqn
R=\frac{9}{2}\frac{B^2r_H^{\ 2}
(5\sin^2{\theta}\sinh^2{\alpha}-2\cos^2{\theta}\cosh^2{\alpha})-6}
{r_H^{\ 5}B^3\sin^3{\theta}\cosh^4{\alpha}}\ . 
\label{curvature}
\eeqn 
Thus, one realizes that the flux brane smoothes out the
irregular horizon, except in the $\theta=0$ three--plane. One
might therefore think that we can travel through the horizon along
the two--plane $\theta=\pi/2$. However, the metric component
$g_{\varphi \varphi}$ for the azimuthal angle in this two--plane
becomes zero at $r=r_H$. This should therefore be faced as the
locus of the origin on this two--plane. Thus, the spacetime is
geodesically complete and there is a singular horizon at $\th=0$,
$r=r_H$. This horizon spans a $\bE^{4}\times S^2$ hypersurface
representing the expansion of the D4--brane into a D6--brane.
Furthermore, the locus $r=r_H$, $0<\th\le \pi/2$, is a regular
three--dimensional surface in the transverse space with the
dielectric two--sphere as a boundary. This is represented in
figure 2.

\begin{figure}
\begin{picture}(0,0)(0,0)
\put(220,135){$r=r_H \equiv {\mathit M}$}
\put(175,182){$_{y_4,y_5}$} \put(176, 169){$_{(\theta=\pi/2)}$}
\put(226,7){$_{(\theta=0)}$} \put(224,18){$_{y_1,y_2,y_3}$}
\put(269,70){$\partial {\mathit M}\equiv S^2$}
\end{picture}
\centering\psfig{file=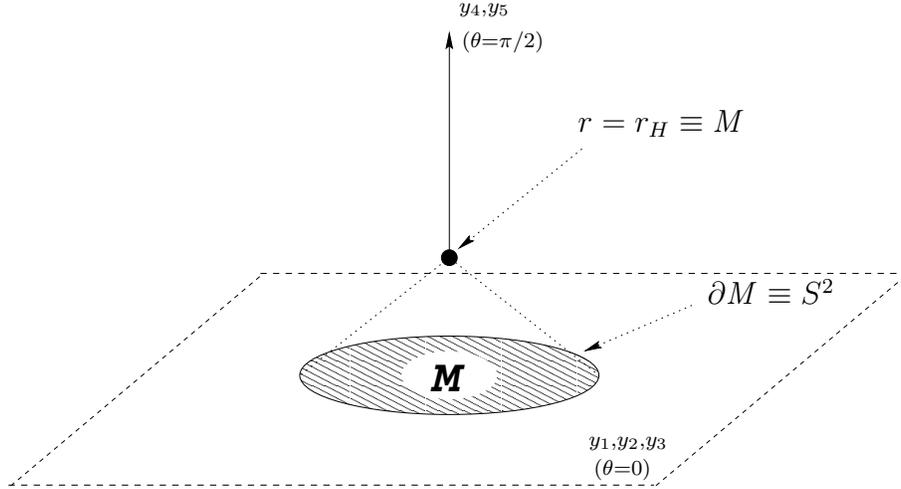,width=12cm}
\caption{\small{Geometry of the hypersurface ${\mathit M}$,
defined as the restriction to the transverse space of the
intersection of $r=r_H$ with a partial Cauchy surface. For $B=0$,
${\mathit M}$ is a point in the transverse space and can be
thought of as the physical singularity associated to the locus of
the D4--branes. As $B$ is turned on, ${\mathit M}$ expands into a
regular three-space bounded by a singular $S^2$, which can be
thought of as the locus of the D4--branes which have expanded into
a D4--D6--bound state. In the next section we will be able to
smoothly interpolate between these two configurations.}}
\label{fig2}
\end{figure}

Before we have a closer look at the geometry near the dielectric
brane, let us discuss the issue of a possible conical singularity
in the $\theta=\pi/2$ two--plane due to the shift in the symmetry
axis introduced in the compactification from M--theory. A simple
analysis on this two--plane shows that for non--vanishing $B$ and
close to $r=r_H$, $\varphi$ would need to be identified with
period 
\eqn 
\frac{4\pi}{3}\,B\cosh{\alpha}\,r_H=2\pi B R_{11}\ .
\label{conical} 
\eeqn 
Since $\varphi$ has periodicity $2\pi$ the
conical singularity is avoided by setting $B=1/R_{11}$. This shows
why we have chosen such value for the magnetic field parameter
earlier.

Now we want to see in more detail that the D4--branes have
expanded to a D6--brane with $N$ D4--branes bound to it and with
worldvolume geometry $\bM^{5}\times S^2$. First consider the usual
solution describing a `flat' bound state of D4--branes within
D6--branes. The bulk fields are 
\eqn 
\arr{c}
\displaystyle{ds_{10}^{\ \ 2}=F^{-1/2}ds^2\left(\bM^{5}\right)+
F^{1/2}\tilde{F}^{-1}ds^2\left(\bE^2\right)+
F^{1/2}ds^2\left(\bE^3\right)}\ ,
\spa{0.3}\\
\displaystyle{e^{2\phi}=F^{-1/2}\tilde{F}^{-1}\ , \ \ \ \ {\cal
A}_1=-\mu\cos{\d}\,\cos{\eta}\,d\varphi}\ ,
\spa{0.3}\\
\displaystyle{{\cal
B}=\tan{\d}\,\tilde{F}^{-1}\e\left(\bE^2\right)\ ,\ \ \ \ {\cal
A}_3=-\mu\sin{\d}\,\tilde{F}^{-1}
\cos{\eta}\,d\varphi\w\e\left(\bE^2\right)}\ . 
\earr 
\label{D4-D6}
\eeqn 
The metric functions read 
\eqn F=1+\frac{\mu}{\rho}\ , \ \ \ \ 
\tilde{F}=\sin^2{\d}+F\cos^2{\delta}\ , 
\eeqn 
and $(\eta,\varphi)$ are the usual angles on a 2--sphere. These fields
are obtained by T--duality at an angle starting from the D5--brane
solution \cite{Brec:96,CostaPapa:96}. Near the `core', i.e. for
$\rho\ll \m$, this solution has the form 
\eqn 
\arr{c}
\displaystyle{ds_{10}^{\ \ 2}=
\sqrt{\frac{\rho}{\mu}}\,ds^2\left(\bM^{5}\right)+
\frac{1}{\cos^2{\d}}\sqrt{\frac{\rho}{\mu}}\,ds^2\left(\bE^2\right)+
\sqrt{\frac{\mu}{\rho}}\,ds^2\left(\bE^3\right)}\ ,
\spa{0.5}\\
\displaystyle{e^{2\phi}=\frac{1}{\cos^2{\d}}
\left(\frac{\rho}{\mu}\right)^{3/2}\ ,\ \ \ \ {\cal
A}_1=-\mu\cos{\d}\,\cos{\eta}\,d\varphi}\ ,
\spa{0.5}\\
\displaystyle{{\cal B}=
\frac{\tan{\d}}{\cos^2{\d}}\,\frac{\rho}{\mu}\,\e\left(\bE^2\right)\
,\ \ \ \ {\cal
A}_3=-\frac{\sin{\d}}{\cos^2{\delta}}\,\rho\,\cos{\eta}\,
d\varphi\w\e\left(\bE^2\right)}\ . 
\earr 
\label{D4-D6core} 
\eeqn
We wish to compare these fields with the ones of solution
(\ref{gravdiele}) near the `core' $r=r_H$, $\theta=0$. It is
convenient to introduce the dimensionless radial coordinate 
\eqn
\la^2=\frac{2}{3}\left(\frac{\tilde{r}}{r_H}\right)^2\ ,
\label{lambda} 
\eeqn 
where $\tilde{r}^2=r^2-r_H^{\ 2}$. Then the
metric in (\ref{gravdiele}) becomes 
\eqn 
\displaystyle{ds_{10}^{\
\ 2}= \frac{3\sqrt{\la^2+\th^2}}{2}\left\{
\frac{ds^2\left(\bb{M}^{5}\right)}{\cosh{\a}}+ r_H^{\
2}\cosh{\a}\left(d\la^2+d\th^2+d\O_2^{\ 2}+
\frac{\th^2\la^2}{\th^2+\la^2}\,d\varphi^2\right)\right\}}\ .
\label{coremetric} 
\eeqn 
The metric in the $\th,\la,\varphi$
three--space is conformally flat. This is made explicit by the
coordinate transformation 
\eqn 
\frac{4\rho}{3r_H}=\la^2+\th^2\ , \
\ \ \ \frac{2\rho}{3r_H}\,\sin{\eta}=\la\,\th\ .
\label{coordtrans} 
\eeqn 
Moreover, with the identifications 
\eqn
\frac{3}{r_H\cosh^2{\a}}\leftrightarrow\frac{1}{\mu}\ , \ \ \ \
\cosh{\a}\leftrightarrow\frac{1}{\cos{\d}}\ , 
\label{ident1} 
\eeqn
we find the `near--core' metric, together with the dilaton and
gauge fields to have the form 
\eqn 
\arr{c}
\displaystyle{ds_{10}^{\ \ 2}=
\sqrt{\frac{\rho}{\mu}}\,ds^2\left(\bM^{5}\right)+
\frac{1}{\cos^2{\d}}\,\sqrt{\frac{\rho}{\mu}}\,r_H^{\ 2}\,d\O_2^{\
2}+ \sqrt{\frac{\mu}{\rho}}\,ds^2\left(\bE^3\right)}\ ,
\spa{0.5}\\
\displaystyle{e^{2\phi}=\frac{1}{\cos^2{\d}}
\left(\frac{\rho}{\mu}\right)^{3/2}\ ,\ \ \ \ {\cal B}=
\frac{\tan{\d}}{2\cos^2{\d}}\,\frac{\rho}{\mu}
\left(1-\cos{\eta}\right)\,r_H^{\ 2}\,\e\left(S^2\right)}\ ,
\spa{0.5}\\
\displaystyle{{\cal A}_1=
\mu\cos{\d}\left(1-\cos{\eta}\right)d\varphi\ ,\ \ \ \ {\cal
A}_3=\frac{\sin{\d}}{\cos^2{\d}}\,\rho
\left(1-\cos{\eta}\right)\,r_H^{\ 2}\,\e\left(S^2\right)\w
d\varphi}\ . 
\earr 
\label{dielcore} 
\eeqn 
Comparison with (\ref{D4-D6core}) reveals that near the `core', the
metric, dilaton field and RR 1--form potential are the same, except
for the crucial replacement of 
$ds^2\left(\bb{E}^2\right)\rightarrow r_H^{\ 2}\,d\Omega_2^{\ 2}$, 
which is exactly what one would
expect from the spherical D4--D6--branes bound state in the
dielectric effect. For this reason it is natural to regard $r_H$
as the radius of the dielectric brane. The 2--form and 3--form
potentials ${\cal B}$ and ${\cal A}_3$ are not in the same gauge
orbit as those in the solution (\ref{D4-D6core}). This fact is
expected because these fields have components along the 2--sphere.
We have checked that both ${\cal B}$ and ${\cal A}_3$ in
(\ref{dielcore}) solve their equations of motion: 
\eqn 
\arr{c}
\displaystyle{d\left(\star\left[e^{-2\phi}{\cal H}- {\cal
A}_1\w\star\tilde{\cal F}_4\right]\right)= -\frac{1}{2}{\cal
F}_4\w{\cal F}_4}\ ,
\spa{0.4}\\
\displaystyle{d\left(\star\tilde{{\cal F}}_4\right)= {\cal
F}_4\w{\cal H}}\ , 
\earr 
\label{formeqns} 
\eeqn 
where
$\tilde{{\cal F}}_4={\cal F}_4-{\cal H}\w{\cal A}_1$.

Since anti--podal points on the 2--sphere have opposite D6--brane
charge, the total charge vanishes. From the source term in the RR
1--form potential equation of motion one finds the usual
quantization of the D6--brane tension: 
\eqn
\mu\,\cos{\d}=\frac{r_H\cosh{\a}}{3}=\frac{g\sqrt{\a'}}{2}N_6\ .
\label{N6} 
\eeqn 
Using the relation (\ref{periodtau}) one sees
that $N_6=1$. Hence, the solution (\ref{gravdiele}) is interpreted
as the ground state of dielectric branes described in terms of
$SU(2)$ irreducible representations, as explained in the
Introduction.

Another consistency check is to view the system as a bound state
of a D6--brane with $N$ D4--branes due to a flux of the $U(1)$
worldvolume gauge field (or gauge equivalently of the Kalb--Ramond
2--form field ${\cal B}$). A simple calculation shows that 
\eqn
\Phi=-\frac{1}{2\pi\a'}\int_{\partial{\cal M}}{\cal B}=2\pi N\ .
\label{flux} 
\eeqn

Finally, the radius of the dielectric sphere $r_H$ is 
\eqn
r_H\sim\frac{\sqrt{\a'}}{g}N=\a' BN\ , \label{radius} 
\eeqn 
in agreement with the scaling behavior in the brane picture
(\ref{myersrad}). We shall return to this point below.

\subsection{Arbitrary magnetic field}

The above construction is not general because the magnetic field
has an unphysically large value $B=1/R_{11}$. We chose to present
it as a warm up to the general case because it is simpler and
encodes the correct physics. To obtain an arbitrary value for the
magnetic field parameter one needs to consider a version of the
rotating M5--brane solution \cite{Cvet:96,Sfet:99} obtained after
a double analytic continuation. The corresponding
eleven--dimensional background fields are 
\eqn 
\arr{c}
\displaystyle{ds^2= H^{-1/3}\left[ds^2\left(\bM^{5}\right)
+f\left(d\t+\frac{2ml\cosh{\a}}{r^3\D f}
\sin^2{\th}\,d\tilde{\varphi}\right)^2\right]+}
\spa{0.6}\\
\displaystyle{+H^{2/3}\left[\frac{dr^2}{\tilde{f}}+
r^2\left(\D\,d\th^2+\sin^2{\th}\,\frac{\D\tilde{f}}{f}\,d\tilde{\varphi}^2
+\cos^2{\th}\,d\O_2^{\ 2}\right)\right]}\ ,
\spa{0.6}\\
\displaystyle{{\cal A}=\frac{2m\sinh{\a}}{\D}\cos^3{\th}
\left[-\left(1-\frac{l^2}{r^2}\right)\cosh{\a}\,d\tilde{\varphi}
-\frac{l}{r^2}\,d\t\right]\w\e\left(S^2\right)}\ . 
\earr 
\eeqn 
The functions $H$, $\D$, $f$ and $\tilde{f}$ have the form 
\eqn
\arr{c} 
\displaystyle{H=1+\frac{2m\sinh^2{\a}}{\D r^3}\ ,\ \ \ \ \
\D=1-\frac{l^2\cos^2{\th}}{r^2}}\ ,
\spa{0.4}\\
\displaystyle{f=1-\frac{2m}{\D r^3}\ ,\ \ \ \ \ \
\tilde{f}=\frac{r^3-l^2r-2m}{r^3\D}}\ . 
\earr 
\eeqn 
This solution
has three independent parameters, namely $(m,\a,l)$, which after
reduction will be related to the number of D4--branes $N$, the
string coupling $g$ and the magnetic field $B$. Notice that this
double analytically continued solution is static. The initial
rotation is mapped into a `twist' along what will become the
internal direction.

The quantization of the M5--brane charge gives the relation 
\eqn
2m\cosh{\a}\sinh{\a}=\pi N l_P^{\ 3}\ . 
\label{M5charge} 
\eeqn 
The `Euclidean angular velocity' and the radius of the compact
direction $\t$ are 
\eqn 
\O=\frac{l}{\cosh{\a}\left(r_H^{\
2}-l^2\right)}\ , 
\label{Omega}
\eeqn 
and 
\eqn
R_{11}=g\sqrt{\a'}=\frac{4 m\cosh{\a}}{3r_H^{\ 2}-l^2}\ ,
\label{R11} 
\eeqn 
respectively. The constant $r_H$ is the maximum
zero of the cubic equation 
\eqn 
r_H^{\ 3}-l^2r_H=2m \ ,
\label{m}
\eeqn 
and corresponds to the locus of the horizon in the
usual Lorentzian version of the solution. Explicitly the location
of $r_H$ can be expressed as 
\eqn
r_H=m^{1/3}\left[\left(1+\sqrt{1-\frac{l^6}{27m^2}}\right)^{1/3}
+\left(1-\sqrt{1-\frac{l^6}{27m^2}}\right)^{1/3}\right]\ ,
\label{rH} \eeqn which for $l^3/m<<1$ is \eqn r_H\simeq
(2m)^{1/3}\left(1+\frac{l^2}{(2m)^{2/3}}\right). 
\label{aproxrH}
\eeqn 
In particular, notice that in the double analytically
continued solution, $r_H$ grows from the static value ($r_H^{\
3}=2m$), whereas it would decrease in the usual Lorentzian
solution. Since at $\theta=\pi/2$ the ergosurface is at $r^3=2m$,
in the analytically continued solution there is no ergoregion
outside the horizon. The Euclidean continuation $l\rightarrow il$,
maps the ergoregion inside the $r=r_H$ surface. Therefore, coming
from infinity, the first metric singularity will be at the zero of
$\tilde{f}$ in the analytically continued solution rather than at
the ergosurface $f=0$ of the usual Lorentzian geometry.

The 4--sphere at $r=r_H$ is the set of fixed points of the Killing
vector field 
\eqn 
k=\frac{\p\ }{\p\t}-\O\frac{\p
}{\p\tilde{\varphi}}\ . \label{killinghor} 
\eeqn 
If we reduce
along this direction the compactified space--time will have a
4--sphere singularity at $r=r_H$. If instead we reduce along 
\eqn
q=\frac{\p\ }{\p\t} +\left(\frac{1}{R_{11}}-\O\right)\frac{\p\
}{\p\tilde{\varphi}}\ , 
\label{redvector} 
\eeqn 
the set of fixed
points will be a 2--sphere at $r=r_H$, $\th=0$. The resulting
solution will describe a D4--brane immersed in a flux 7--brane
with magnetic field parameter given by 
\eqn
B=\frac{1}{R_{11}}-\O=\frac{3r_H+l}{2r_H(r_H+l)\cosh{\a}}\ ,
\label{generalB} 
\eeqn 
where we have used the previous relations
for $\Omega$ and $R_{11}$ together with the cubic equation for
$r_H$. Defining the azimuthal coordinate
$\varphi=\tilde{\varphi}-B\t$, the eleventh direction $\t$ will
have periodicity $ R_{11}= g\sqrt{\a'}$ at fixed $\varphi$. Then,
the reduction to the type IIA theory yields the following fields:
\eqn 
\arr{c} 
\displaystyle{ds_{10}^{\ \ 2}=
\left(\frac{\Sigma}{H}\right)^{1/2} ds^2\left(\bM^{5}\right)+
\left(\Sigma H\right)^{1/2}\left[\frac{dr^2}{\tilde{f}}+
r^2\left(\D\,d\th^2+\cos^2{\th}\,d\O_2^{\ 2}\right)\right]+}
\spa{0.5}\\
\displaystyle{+\left(\frac{H}{\Sigma}\right)^{1/2}
\tilde{f}\,r^2\sin^2{\th}\,\D\,d\varphi^2}\ ,
\spa{0.5}\\
\displaystyle{e^{2\phi}=\frac{\Sigma^{3/2}}{H^{1/2}}}\ ,\ \ \ \
\displaystyle{{\cal B}=-\frac{2m\sinh{\a}\,\cos^3{\theta}}{\D}
\left[\frac{l}{r^2}+
\left(1-\frac{l^2}{r^2}\right)B\cosh{\alpha}\right]
\epsilon\left(S^2\right)}\ ,
\spa{0.6}\\
\displaystyle{{\cal A}_1=\Psi\,\Sigma^{-1}d\varphi}\ , \ \ \ \ \
\displaystyle{{\cal A}_3=
-\frac{2m\sinh{\a}\,\cosh{\a}\,\cos^3{\theta}}{\D}
\left(1-\frac{l^2}{r^2}\right)d\varphi\w
\epsilon\left(S^2\right)}\ , 
\earr 
\label{gravdieleB} 
\eeqn 
where we have introduced 
\eqn 
\Sigma\equiv
f\left(1+\frac{2mlB\cosh{\a}\, \sin^2{\th}}{r^3\D f}\right)^2+
H(Br\sin{\th})^2\,\D\,\frac{\tilde{f}}{f}\ , 
\label{SigmaB} 
\eeqn
and 
\eqn 
\Psi\equiv\frac{\Sigma-f}{B}-
\frac{2ml\cosh{\a}\,\sin^2{\th}}{r^3\D}\ . 
\label{Psi} 
\eeqn

Again, this solution describes a geometry with 5--dimensional
Poincar\'e invariance. The number of D4--branes is 
\eqn
N=\frac{1}{(2\pi)^3g\a'^{3/2}}\int {\cal F}_4=
\frac{\sinh{\a}}{2\pi \a'}\,\left(3r_H^{\ 2}-l^2\right) \ .
\label{NB} 
\eeqn 
Also, asymptotically one recovers the flux
7--brane solution (\ref{Melvin}).

\subsubsection{Analysis of Parameters}

As pointed out in the discussion above, the supergravity solution
depends on three parameters. It is convenient to choose as
independent parameters $r_{H}$, $l$ and $\alpha $, with $m$
implicitly defined by equation (\ref{m}). Positivity of the
parameter $m$ requires $r_{H}>\left| l\right| $. In terms of these
parameters the physical quantities $g$, $B$ and $N$ are  defined
by the equations (\ref{R11}), (\ref{generalB}) and (\ref{NB}),
respectively. We have already seen that the $l=0$ case corresponds
to an unphysically large magnetic field $B=1/R_{11}$. The physical
regime where $B$ can be tuned to an arbitrary small value,
$B<<1/R_{11}$, corresponds to the case $r_{H}-l\ll r_{H}$. In this
limit, the physical parameters become 
\eqn 
\arr{rcl}
N &\simeq &
\displaystyle{\frac{r_{H}^{\ 2}\sinh{\a}}{\pi\a'}}\ , 
\spa{0.5}\\
B &\simeq &
\displaystyle{\frac{1}{r_{H}\cosh{\a}}} \ , \spa{0.6}\\
g\sqrt{\a'} &\simeq &
\displaystyle{2\left(r_{H}-l\right)\cosh{\a}}\ . 
\earr
\label{NBg}
\eeqn

A careful analysis of the parameters shows that, for fixed $g$ and
$B$, the number of $D4$--branes $N$ is bounded above by a critical 
number $N_{\mathrm{crit}}\left( g,B\right)$. 
In the limit just described, this can be shown simply by noting that
\begin{equation}
\pi\a'NB^{2}\simeq \frac{\sinh{\a}}{1+\sinh^{2}{\a}}
\leq \frac{1}{2}\ ,
\end{equation}
so that
\begin{equation}
N_{\mathrm{crit}}\simeq \frac{1}{2\pi \alpha ^{\prime }B^{2}}\ .
\end{equation}
Similarly, for fixed $g$ and $N$ there is a critical magnetic field 
$B_{\rm crit}\simeq 1/\sqrt{2\pi\a'N}$. Notice that the existence of
an upper bound on the D4--brane charge is another manifestation of the
stringy exclusion principle \cite{MaldStro}.

Finally, let us note that the map $\left( r_{H},l,\alpha \right)
\rightarrow \left( g,N,B\right) $ is generically two--to--one,
except at the special points where the parameters $N,B$ acquire
their critical value. In particular for a given set $(g,N,B)$
there will be generically two values of the dielectric sphere
radius $r_H$. The physical interpretation of this fact will
follow.\footnote{Before the revised version of this paper came out, a
pre--print \cite{BrecSaff} appeared that independently realized, from
the analysis of the parameters of our solution, the existence of two
radii and of a critical number of branes.}

\subsubsection{Pure Melvin and Pure D4--brane Limits}

Before we consider the general case, let us analyse the geometry
for $l=r_H$. Then the parameter $m$ vanishes and we recover the
flux 7--brane solution written in spheroidal oblate coordinates,
which can be related to the usual cartesian coordinates by the
transformation 
\eqn 
\arr{c} 
\left( 
\arr{c} 
y_1
\\ y_2+iy_3
\\ y_4+iy_5
\earr 
\right)=\left( \arr{c} r\cos{\theta}\,\cos{\psi}
\\ r\cos{\theta}\,\sin{\psi}\,e^{i\chi}
\\ \sqrt{r^2-r_H^{\ 2}}\,\sin{\theta}\,e^{i\varphi}
\earr 
\label{spheroidal} 
\right)\ . 
\earr 
\eeqn 
To recover flat space one should send $\a\rightarrow\infty$, since from
(\ref{generalB}) the magnetic field $B$ vanishes in such limit.

Next, we consider the limit of vanishing magnetic field keeping
the total number of D4--branes $N$ and the string coupling $g$
fixed. For $B$ small enough we will always be in the region of
parameter space $r_H-l\ll r_H$ defined in the previous subsection.
More precisely inverting the relations (\ref{NBg}) we have 
\eqn
\arr{rcl}
r_H &\simeq &
\displaystyle{\pi\a' NB\rightarrow 0} \ , \spa{0.4}\\
\displaystyle{\frac{r_H-l}{r_H}} &\simeq &
\displaystyle{\frac{1}{2}\,g\sqrt{\a '}B\rightarrow 0}\ , \spa{0.4}\\
\cosh{\a}&\simeq &
\displaystyle{\frac{1}{\pi\a' N B^2}\rightarrow
\infty }\ .
\earr
\label{invertNBg}
\eeqn 
In this limit $f\rightarrow 1$ and the forms
${\cal A}_1$ and ${\cal B}$ in  (\ref{gravdieleB}) vanish. The
resulting solution is that of the single--center D4--brane
geometry, as expected since for vanishing magnetic field the
D4--branes will seat on top of each other. Conversely, starting
with the single--center D4--brane geometry, which is a special
case of ({\ref{gravdieleB}), we can turn on a magnetic field and
observe the expansion of the system into a spherical dielectric
brane.

\subsubsection{General Case}

Similarly to the maximal magnetic field case, the geometry
described by (\ref{gravdieleB}) has a metric singularity at
$r=r_H$. The geometry has exactly the same features as described
by Figure 2. The locus $r=r_H$ is the origin on the $\th=\pi/2$
two-plane, and the D6--brane worldvolume with geometry
$\bM^{5}\times S^2$ is represented by $r=r_H$ and $\th=0$. The
three-dimensional space $r=r_H$, $0<\th\le \pi/2$ in the
transverse space represents the interior of the dielectric
two--sphere.

The analysis of the geometry near $r=r_H$, $\th=\pi/2$ shows that
to avoid a conical singularity the following condition must hold
\eqn 
3m+l^2r_H=lr_H^{\ 2}+2mr_H B\cosh{\a}\ . 
\label{conicalB}
\eeqn 
Using the cubic equation for $r_H$, this yields relation
(\ref{generalB}) and explains why we chose this particular value
for the magnetic field parameter.

Now we show that the geometry near $r=r_H$, $\th=0$ describes a
spherical D6--D4--brane bound state as appropriate for the
dielectric effect. In this limit the functions appearing in the
metric become 
\eqn 
\arr{c} 
\displaystyle{H\simeq \cosh^2{\alpha}\ , 
\ \ \ \ \ \ 
\Delta\simeq \frac{2m}{r_H^{\ 3}}+\frac{l^2}{r_H^{\ 2}}
\left(\theta^2+\frac{\tilde{r}^2}{r_H^{\ 2}}\right)}\ ,
\spa{0.6}\\
\displaystyle{f\simeq \frac{3r_H^{\ 2}-l^2}{4mr_H}\,\tilde{r}^2
+\frac{l^2r_H}{2m}\,\theta^2\ , 
\ \ \ \ \ \ 
\tilde{f}\simeq\frac{3r_H^{\ 2}-l^2}{4mr_H}\,\tilde{r}^2}\ , 
\earr
\label{metricHDfs} 
\eeqn 
where as before we defined the radial
coordinate $\tilde{r}^2=r^2-r_H^{\ 2}$. It is convenient to change
to the new coordinates 
\eqn
\rho=\frac{\tilde{r}^2}{2r_H}+\frac{3r_H^{\ 2}-l^2}{4r_H}\,\th^2\
, \ \ \ \ \rho\sin{\eta}=\sqrt{\frac{3r_H^{\ 2}-l^2}{2r_H^{\
2}}}\,\tilde{r}\,\th\ . 
\label{coortrans} 
\eeqn 
With the identification 
\eqn 
\mu \leftrightarrow
\frac{2m\cosh^2{\alpha}}{3r_H^{\ 2}- l^2}\ , \ \ \ \
\cosh{\alpha}\leftrightarrow \frac{1}{\cos{\delta}}\ ,
\label{ident2} 
\eeqn 
we obtained exactly the same metric, dilaton
field and RR 1--form and 3--form potentials as in the `near--core'
geometry for the maximal magnetic field case (\ref{dielcore}). The
2--form Kalb-Ramond field is now 
\eqn 
{\cal B}=\frac{\tan{\delta}}{\cos^2{\delta}}
\left(\frac{1}{\mu}\,\rho\sin^2{(\eta/2)}
+\frac{lr_H\cos^2{\delta}}{m}\,\rho\cos^2{(\eta/2)}\right) r_H^{\
2}\,\e(S^2)\ . 
\label{2form} 
\eeqn 
As before we verified that the
equations of motion (\ref{formeqns}) for the 2 and 3--form are
obeyed for this `near--core' solution. Again, therefore, the
interpretation of a spherical bound state holds, as appropriate
for the dielectric effect.

The quantization condition on the D6-brane tension gives 
\eqn
\m\,\cos{\d}=\frac{2m\,\cosh{\a}}{3r_H^{\ 2}-l^2}=
\frac{g\sqrt{\a'}}{2}\ , 
\label{D6quant} 
\eeqn 
which together with
the equation for $R_{11}=g\sqrt{\a'}$ in (\ref{R11}) gives
$N_6=1$. Hence, as for the maximal magnetic field we consider the
case with a single D6--brane. Viewing the system as a bound state
of a D6--brane with $N$ D4--branes requires a flux quantization of
the Kalb--Ramond 2--form field ${\cal B}$ along the two-sphere.
Indeed a straightforward calculation yields the same result as in
(\ref{flux}).

The region of validity for the Melvin background requires that
$R_{11}\ll r\ll 1/B$. The former condition arises by considering
length scales much larger than the compactification scale and the
latter by requiring small string coupling. Let us again take $g$
and $N$ fixed, and vary $B$. As we already noted in section 3.2.1,
$0<B<B_{\mathrm{crit}}$. To work within the region of parameter
space $r_{H}-l\ll r_{H}$, we will require that 
$B_{\mathrm{crit}}\ll 1/R_{11}$. Since in this case 
$B_{\mathrm{crit}}\simeq 1/\sqrt{2\pi \alpha ^{\prime }N}$, 
this requirement is equivalent to the mild condition $\sqrt{N}\gg g$.

Now we can calculate the radius of the dielectric brane using the
gravitational approximation. Using the relations (\ref{NBg}) for
$B$ and $N$  in terms of $r_H$ and $\a$ in the limit
$r_{H}-l\ll r_{H}$, we find the quartic equation for $r_H$ 
\eqn
r_{H}^{\ 4}-\left(\frac{r_{H}}{B}\right)^2+\left( \pi \alpha
^{\prime }N\right) ^{2} =0 \ .  
\eeqn 
As mentioned before one
finds two solutions for $r_H$ in terms of the physical parameters,
\eqn 
r_{H}^{\ 2} = \frac{1}{2B^{2}}\left( 1\pm \sqrt{1-\left( 2\pi
\alpha ^{\prime }NB^{2}\right) ^{2}}\right) \ . 
\eeqn 
For small
magnetic field parameter, $B<<B_{\mathrm{crit}}$, we obtain 
\eqn
r_-= \pi \alpha ^{\prime }NB\ ,
\ \ \ \ \ \ 
r_+ =\frac{1}{B}\ .
\label{radii}
\eeqn 
We shall come back to these results in section 5. Here we shall
mention just that $r_H=r_-$ is exactly the radius found by Myers in
his approach, and that $r_H=r_+$ is not a stable solution.

\subsubsection{M-theory Orbifold and Multiple D6--branes}

A small extension of the previous construction gives rise to a
dielectric brane which can be regarded as a bound state with $N_6$
D6--branes. The only modification is to consider an orbifold
compactification of M-theory. More precisely, we mod out the 
$\tau-r$ plane by a discrete $\bZ_{N_6}$ identification. This is
implemented by modifying the eleven--dimensional period via 
\eqn
R_{11}=g\sqrt{\a'}=
\frac{4m\cosh{\a}}{3r_H^{ \ 2}-l^2}\, \frac{1}{N_6}\ ,
\eeqn 
which replaces the previous relation (\ref{R11}). The
magnetic field parameter is still related to  $(r_H, l, \a)$ by
(\ref{generalB}), while the total number of D4--branes becomes
\eqn 
N= \frac{N_6}{2\pi \a'}\,\left(3r_H^{\
2}-l^2\right)\sinh{\a}\ . 
\eeqn 
In fact, given a ten--dimensional
solution with parameters $(r_H,l,\a)$ only the ratio $N/N_6$ is
determined. Only in eleven dimensions the individual values of $N$
and $N_6$ are singled out.

To see that $N_6$ should be interpreted as the number of
D6--branes we look at the geometry close to the dielectric sphere.
The associated D6--brane tension is 
\eqn 
\mu \,
\cos{\d}=\frac{2m\,\cosh{\a}}{3r_H^{\ 2}-l^2}=
\frac{g\sqrt{\a'}}{2}\, N_6\ ,
\eeqn 
which, as claimed, should be identified with a dielectric
configuration with multiple D6--branes.

The analysis of the radius $r_H$ follows the one at the end of the
previous section with the replacement $N\rightarrow N/N_6$. In
particular, the dielectric radius $r_-=\pi\a'BN/N_6$ is in
agreement with the brane picture in terms of reducible $SU(2)$
representations (formed from $N_6$ irreducible representations each of 
integer dimension $N/N_6$) \cite{Myers:99}. 
Let us remark that the maximal case
$N_6=N$ is described by the same ten-dimensional geometry as the case
$N=1$, and hence outside the scope of the gravitational description.

For simplicity, in the remainder of this paper we shall restrict
ourselves to the single D6--brane case.

\subsubsection{Gravitational Energy}

Having placed the D4--branes in the magnetic field background
associated with the flux 7--brane it is natural to ask what is the
gravitational energy associated with such configuration. Thus we
are led to compute the spacetime energy. In order to get a finite
result we will use the `reference background subtraction' method
\cite{Hawking:95}, with the flux 7--brane as the reference
background. This method yields the right result for the Ernst
black hole in the Melvin Universe \cite{Ernst}, a situation
somewhat analogous to, although much simpler than, the present
case. The mass calculated using this method should be interpreted
as the dielectric brane mass placed in a flux 7--brane. This mass
is associated with the D4--branes mass together with the
interacting energy with the flux 7--brane.

For the dielectric brane geometry described by (\ref{gravdieleB})
the relevant energy expression is 
\eqn 
{\cal E}=-\frac{1}{8\pi G_{10}}
\left[\int_{S^{\infty}_{t}}\sqrt{h}{\mathcal{N}\mathcal{K}}d^8x
-\int_{S^{\infty}_{t}}\sqrt{h_0}{\mathcal{N}}_0{\mathcal{K}}_0d^8x
\right]\ . 
\label{energyfor} 
\eeqn 
Here, we have denoted by
$h_{\alpha \beta}$ the induced metric on a co-dimension two
surface which can be thought of as a $r=const.$ (as $r\rightarrow
\infty$) section of a spacelike hypersurface. $\mathcal{K}$ is the
trace of the second fundamental form on $S^{\infty}_{t}$ and
${\mathcal{N}}=\sqrt{-g_{00}}$ is the lapse function. The
quantities with a zero subscript refer to the flux 7--brane
background. Explicitly the trace of the second fundamental form
reads 
\eqn 
{\mathcal{K}}=
\frac{1}{2\sqrt{g_{rr}}}\left(h^{\a\b}h_{\a\b,r}\right)\ .
\label{K} \eeqn Then a straightforward computation leads to the
result \eqn {\cal E}=\frac{\pi V_4}{3G_{10}}
\left(\frac{3}{2}R^3+5m\right)\ , 
\label{energy} 
\eeqn 
where $V_4$
stands for the volume of ${\bb E}^4$ and we used the value for the
unit 4--sphere volume $\Omega_4=8\pi^2/3$.

To relate this expression to the D4--branes mass plus interaction
with the flux 7--brane, consider the limit of small magnetic field
defined in section 3.2.1. Writing $R^3$ and $2m$ as
\eqn
\arr{c}
\displaystyle{R^3=2m\sinh^2{\a}=2m\left(\cosh^2{\a}-1\right)}\ ,
\spa{0.3}\\
\displaystyle{2m=r_H\left(r_H^{\ 2}-l^2\right)
\simeq 2r_H^{\ 3}\,\frac{r_H-l}{r_H}}\ ,
\earr
\eeqn
and using the relations (\ref{invertNBg}), together with the formulae
for the gravitational constant and D--brane tension
\eqn 
16\pi G_{10}=(2\pi)^7g^2\alpha'^4\ , 
\ \ \ \ \
T_{p}=\frac{1}{(2\pi)^p g\, \alpha'^{(p+1)/2}}\ , 
\label{GT} 
\eeqn
we arrive at the result
\eqn 
{\cal E}\simeq NM_{4}+\frac{2\pi^2\alpha'^2}{3}B^4N^3 M_{4}\ . 
\label{E} 
\eeqn
Here $M_4=V_4T_4$ is the mass of one D4-brane in flat space. The second
term should be identified with the interaction between the
D4-branes and the flux 7--brane that results in the dielectric
configuration. One could think that this energy is higher than the
mass of $N$ D4--branes and therefore the solution is unstable.
Notice, however, that the vacuum, i.e. the reference background,
is not flat space but the flux 7--brane which does not allow for a
configuration of $N$ D4--branes placed on top of each other with
the corresponding flat space mass.

\section{Decoupling Limit}
\news

The gravity/gauge theory duality holds in the limit when closed strings in
the bulk and open strings on the D--brane decouple \cite{Mald:97}. On the
gauge theory side we expect that the coupling of the D4--branes to the flux
7--brane will generate a relevant deformation on the D4--brane worldvolume
theory, controlled by the magnetic field $B$. Of course, the most
interesting case would be the one involving D3--branes expanding to a
spherical D5--brane, which is analogous to the one presented in this paper.

The gravitational dual is obtained by taking the decoupling limit on the
geometry (\ref{gravdieleB}). This limit corresponds to sending 
$\a'\rightarrow 0$ keeping fixed the parameters 
\eqn
g_{YM}^{2}=(2\pi )^{2}g\sqrt{\alpha ^{\prime }}\ ,
\ \ \ \ \ \ \ \ \ \ \ \
U_{0}=\frac{r_{H}}{\alpha ^{\prime }}\ ,
\ \ \ \ \ \ \ \ \ \ \ \ \ \ \ \ 
a=\frac{l}{\alpha ^{\prime }}\ ,  
\label{declim}
\eeqn
and the energy scale 
\eqn
U=\frac{r}{\alpha ^{\prime }}\ .
\eeqn
In this limit, the extremality parameter $\a\rightarrow \infty$ so
that $\a'\cosh{\a}$ remains finite with value 
\eqn
\kappa =\frac{g_{YM}^{2}}{8\pi ^{2}}
\left( \frac{3U_{0}^{\ 2}-a^{2}}{U_{0}^{\ 3}-a^{2}U_{0}}\right).
\label{kappa}
\eeqn
Moreover, the magnetic field parameter $B$ is kept fixed, and can be
expressed as
\begin{equation}
B=\frac{\left( 2\pi \right) ^{2}}{g_{YM}^{2}}\frac{\left(
3U_{0}+a\right) \left( U_{0}-a\right) }{3U_{0}{}^{2}-a^{2}\,}\ .
\label{Bdecoupl}
\end{equation}
A straightforward calculation shows that the background fields have
the form 
\eqn 
\arr{c} 
\displaystyle{ ds_{10}^{\ \ 2}=\a'\left\{
\left(\frac{\Sigma}{H}\right)^{1/2}ds^2\left(\bM^{5}\right)+
\left(\Sigma H\right)^{1/2}\left[\frac{dU^2}{\tilde{f}}+
U^2\left(\D\,d\th^2+\cos^2{\th}\,d\O_2^{\ 2}\right)\right]+\right.} 
\spa{0.6}\\
\displaystyle{\left.\phantom{\frac{dU}{\tilde{f}}}
+\left(\frac{H}{\Sigma}\right)^{1/2}
\tilde{f}\,U^2\sin^2{\th}\,\D\,d\varphi^2\right\}}\ ,
\spa{0.5}\\
\displaystyle{g^2e^{2\phi}=
\frac{g^4_{YM}}{(2\pi)^4}\,\Sigma^{3/2}H^{-1/2}\ ,
\ \ \ \ \ \ \ \ \ \ 
{\cal A}_1=\Psi\,\Sigma^{-1}d\varphi}\ ,
\spa{0.7}\\
\displaystyle{{\cal B}=
-\a'\,\frac{g^2_{YM}N}{4\pi}\,\frac{\cos^3{\th}}{\D}
\left[B+\frac{a-\kappa Ba^{2}}{\kappa }\frac{1}{U^{2}}\right]
\e\left(S^2\right)}\ ,
\spa{0.7}\\
\displaystyle{{\cal A}_3=
-\a'\,\frac{g^2_{YM}N}{4\pi}\,\frac{\cos^3{\th}}{\D}
\left(1-\frac{a^2}{U^2}\right)d\varphi\w\e\left(S^2\right)}\ ,
\earr 
\label{decoupling} 
\eeqn
where the functions $H$, $\Sigma$ and $\Psi$ now read 
\eqn
\arr{c}
\displaystyle{H=\frac{g_{YM}^{2}N}{4\pi \Delta U^{3}}}\ ,
\spa{0.5}\\ 
\displaystyle{\Sigma =
f\left( 1+\frac{g_{YM}^{2}aBN}{4\pi \kappa }
\frac{\sin^{2}{\theta }}{\Delta fU^{3}}\right)^{2}
+\frac{g_{YM}^{2}NB^{2}}{4\pi }\,\frac{\sin ^{2}{\theta }}{U}
\frac{\tilde{f}}{f}}\ ,
\spa{0.6}\\
\displaystyle{\Psi =
\frac{\Sigma -f}{B}-\frac{g_{YM}^{2}aN}{4\pi \kappa }\,
\frac{\sin ^{2}{\theta }}{\Delta U^{3}}}\ ,
\earr
\label{decSigmaPsi}
\eeqn
with 
\eqn
\Delta =1-\frac{a^{2}\cos ^{2}{\theta }}{U^{2}}\ ,
\ \ \ \ \ \ \ \ 
f=1-\frac{2m_{0}}{\Delta U^{3}}\ ,
\ \ \ \ \ \ \ 
\tilde{f}=\frac{U^{3}-a^{2}U-2m_{0}}{\Delta U^{3}}\ .  
\label{decfunc}
\eeqn
The constant $m_{0}$ is defined through the cubic equation
$2m_{0}=U_{0}^{\ 3}-a^{2}U_{0}$.

Consider first the asymptotics of the above solution. For 
\eqn
U\gg U_{0},\,\ g_{YM}^{2}NB^{2}
\label{asympt}
\eeqn
(recall that $U_{0}>\left| a\right| ,m_{0}^{1/3}$) it is easy to see that
one obtains the same asymptotics as for the decoupling limit of the
D4--brane geometry defined in \cite{Itzh:98}. This fact supports the result
announced at the end of section two: if one places $N$ D$p$--branes in a
flux $(p+3)$--brane and takes the decoupling limit of \cite{Itzh:98} keeping
the magnetic field fixed, the resulting geometry will
have the same asymptotics as for the decoupled D$p$--brane geometry. From
the field theory point of view, this means that the couplings of the
D--branes to the flux--branes -- {\it e.g.} the Myers coupling -- become
irrelevant in the UV. This fact was used in the analysis of an associated
class of ${\cal N}=1$ supersymmetric $4$--dimensional theories by
Polchinski and Strassler in \cite{PolcStra:00}.

\begin{figure}[t]
\begin{picture}(0,0)(0,0)
\put(30,12){$_{\th}$}
\put(50,48){$_{U=U_0}$}
\put(-6,218){$\r$} 
\put(221,-8){$v$} 
\put(43,-8){\footnotesize{P}} 
\put(-6,40){\footnotesize{Q}} 
\put(65,96){IIA SUGRA}
\put(73,13){Large curvature}
\put(134,174){M-theory region}
\put(98,207){Asymptotic region}
\end{picture}
\centering\psfig{file=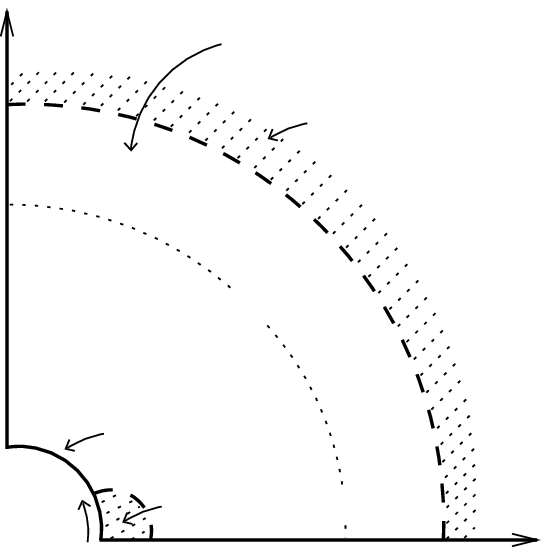,width=8cm}
\caption{\small{The region of validity of the gravitational
approximation, in terms of the variables $v=U\cos{\th}/(2\pi)$ and 
$\r=U\sin{\th}/(2\pi)$. There are curvature corrections
around $U=U_0$ and $\th=0$ (point P), where the dielectric brane is
placed. Inside the dielectric sphere, i.e. for $U=U_0$ and 
$0<\th\le\pi/2$, there is a large region surrounding the center of the
sphere (point Q at $\th=\pi/2$) where the curvature is small.
Far away the appropriate description is eleven--dimensional.}}
\label{fig3} 
\end{figure}

Next we want to analyze the region of validity of the type IIA supergravity
description. We will work in what follows with $g_{YM}^{2}B\ll 1$, which, as
we shall see, is the relevant regime to have a useful supergravity solution
in the large $N$ limit. In this case 
\begin{equation} 
U_{0}\simeq \pi BN\ ,
\  \ \ \ \ \ \ \ \ \ \
\frac{U_{0}-a}{U_{0}}\simeq \frac{g_{YM}^{2}B}{8\pi ^{2}}\ ,
\label{U0,a}
\end{equation}
and condition (\ref{asympt}) simply becomes 
\eqn
U\gg U_{0}\ . 
\eeqn
First we recall some facts from \cite{Itzh:98} on the gravitational
description of the pure D4--brane near horizon geometry. The relevant
quantity is the effective dimensionless 't Hooft coupling $\lambda
_{eff}\left( U\right) =g_{YM}^{2}NU$, in terms of which the region of
validity of supergravity is 
\eqn
1\ll \lambda _{eff}\ll N^{4/3}\ . 
\eeqn
The bound $1\ll \lambda _{eff}$ comes from the requirement of small
curvature, whereas small string coupling gives the bound 
$\lambda _{eff}\ll N^{4/3}$. We want the asymptotic region 
$U\gg U_{0}$ described above to be
still within the SUGRA regime, and for this to hold we simply require
that $\lambda _{eff}\left( U_{0}\right) \ll N^{4/3}$ or 
\eqn
g_{YM}^{2}B\ll N^{-2/3}\ . 
\eeqn
When the above bound holds, as we move away from the dielectric brane, we
reach the asymptotic region {\it before} we arrive at the region of large
string coupling, where Type IIA supergravity ceases to hold and where
we enter the eleven--dimensional M--theory region (see Figure 3). As we
move in towards the dielectric brane to lower values of $U$, we
eventually reach a region of
large curvature, where $\alpha ^{\prime }$ corrections to supergravity are
important. This region is localized around the spherical D6 brane, which, in
the $U$--$\theta $ plane, is located at $U=U_{0}$ and $\theta =0$ (point P
in Figure 3). To analyze the extent of the region of large curvature, we
focus our attention, in particular, on the region inside the dielectric
sphere, around $U=U_{0}$ and $\theta =\pi /2$ (point Q in the
Figure). This point is separated from the spherical brane by a
distance (in energy units) of order $U_{0}$, and therefore the
curvature is small if the effective coupling satisfies 
$\lambda _{eff}\left( U_{0}\right) \gg 1$, or 
\eqn
g_{YM}^{2}B\gg N^{-2}\ . 
\eeqn
When the above holds, the region of large curvature around point P does not
extend all the way to the interior of the dielectric sphere to
point Q and we are in the situation described in Figure 3. This last
bound was obtained with a heuristic argument, but it can be confirmed
by computing the curvature at point Q and requiring it to be small in
units of $\a'$.

\subsection{Scalars VEV and vacuum energy}

The reason for the above analysis is to justify the validity of the
gravitational calculation of both the radius and potential energy of the
dielectric brane in the decoupling limit. These are related to the VEV for
the scalars of the field theory and the energy of the vacuum. Notice
that, however, as it stands our field theory is not renormalizable and
should be regarded as a low energy effective description of a deformation of
the compactified $(2,0)$ six--dimensional conformal theory associated to the
M5--branes.

First recall that the unperturbed theory is free in the IR, and that the
scalars have vanishing expectation value in the vacuum. The deformation,
proportional to $B$, is on the other hand IR relevant, and changes the
vacuum structure. More precisely, the VEV for the scalars is given by 
\eqn
U_{0}=2\pi \langle {\bf \Phi }\cdot {\bf \Phi }\rangle ^{1/2}\ .  
\label{VEVeq}
\eeqn
This is defined implicitly in terms of $B$, $N$ and
$g_{YM}^{2}$. Provided $U_{0}$ is much larger than the thin layer
around the dielectric brane, where curvature corrections are
important, the gravitational approximation is accurate. Now we
consider the limit of small $B$ defined above. Then to
linear order in $B$ we have 
\eqn
U_{0}\simeq \pi BN\ .  
\label{vev}
\eeqn
which exactly corresponds to the Myers result, including the numerical
factor. This value corresponds to the radius $r_{H}=r_{-}$ of section
3 after taking the decoupling limit. The other value $r_{H}=r_{+}$ in
(\ref{radii}) scales to infinity in the decoupling limit.

The calculation of the vacuum energy uses the `reference background
subtraction' method explained in section 3.2.5. Now, the appropriate
reference vacuum is the D4--brane geometry in the decoupling limit. A
straightforward calculation gives 
\begin{equation}
{\cal E}=\frac{4V_{4}m_{0}}{3\pi g_{YM}^{4}}\ .  \label{vacene}
\end{equation}
For small magnetic field we have, to leading order in $B$, 
\begin{equation}
{\cal E}\simeq \frac{V_{4}}{6g_{YM}^{2}}B^{4}N^{3}\ ,  \label{vacene2}
\end{equation}
which is exactly the same energy as the interacting energy term in 
(\ref{E}) after taking the decoupling limit. This vacuum
energy has the same scaling behavior as the vacuum energy $V_{N}$ calculated
by Myers in the flat space case (\ref{myerspot}). Let us note, however, that
the above result for the energy includes, from the field theory point of
view, two contributions. The first comes from the change in the vacuum
energy coming from the new relevant operators proportional to $B$ (the
graphs with {\it no} external legs). The second contribution comes from the
graphs with external legs at zero momentum, since the vacuum value of the
scalar fields is not zero in the dielectric brane configuration.

To better interpret these results notice first that we are considering only
leading terms in $B$ (even if our expressions for $U_{0}$ and ${\cal E}$ are
valid to all orders). The 3--point vertex deformation in the Lagrangian that
is linear in $B$ has the form of (\ref{potential}) and it is the Myers
coupling of the D4--brane to the electric field on the flux brane. The
deformation, as argued before, is relevant in the IR, and therefore graphs
of all orders will contribute to deep IR problems like vacuum structure
(recall that the unperturbed theory becomes free in the IR, since the
effective coupling $\lambda _{eff}\left( U\right) $ goes to zero as
$U\rightarrow 0$). We are, on the other hand, in a position in which we can
fine--tune the magnetic field to an arbitrarily small value, and therefore
we can see results in perturbation theory in $B$ regarding IR physics. We
therefore conclude that (\ref{VEVeq}) and (\ref{vacene}) are strong coupling
(in $B$) results for the vacuum structure of the perturbed theory, given a
small deformation in the UV. We will say a bit more on the form of the
high--energy Lagrangian and deformation in the next section, using a probe
computation in the decoupled geometry. It would be very interesting to use
these gravity predictions to match some field theory results.

\section{Probing Dielectric Branes}

In this section we will study the flux--brane and dielectric brane
geometries using a D--brane probe. Let us start by discussing a
$D6$--brane probe in the presence of a flux $7$--brane. Following
Myers \cite{Myers:99}, we will take the probe geometry to be
$\bM^{5}\times S_{2}$, with $n$ units of $D4$--brane charge
arising from a constant $U\left( 1\right) $ flux on the
two--sphere. The action for a $D6$--brane probe is
\begin{equation}
S=S_{BI}+S_{WZ}\ ,
\end{equation}
where
\eqn
\arr{c}
\displaystyle{S_{BI}=-T_{6}\int d^{7}\sigma \,e^{-\Phi}
\sqrt{-\det \left( \hat{g}-B+2\pi\alpha ^{\prime }F\right) }}\ , 
\spa{0.4}\\
\displaystyle{S_{WZ}=T_{6}\int \sum {\mathcal{A}}_{p+1}
\wedge e^{-B+2\pi\alpha ^{\prime }F}}\ .
\earr
\eeqn
The worldvolume magnetic field $F$ has the form
$-\frac{1}{2}n\,\epsilon \left( S_{2}\right) $, and the position
of the probe is parameterized by the two coordinates $r$ and
$\theta $, together with the azimuthal coordinate $ \phi $ which,
by rotational symmetry, does not enter in the following
expressions for the probe potential. Using the flux $7$--brane
background (\ref{Melvin}), one arrives at the potential
\begin{equation}
V=4\pi T_{6}V_{4}\left[ \Lambda ^{1/2} \sqrt{\Lambda \left( r\cos
\theta \right) ^{4}+\left( \pi \alpha ^{\prime }n\right)
^{2}}-\frac{2}{3}B\left( r\cos \theta \right) ^{3}\right] \ ,
\label{add-100}
\end{equation}
with $\Lambda =1+\left( Br\sin \theta \right) ^{2}$. At $\theta
=0$, the above potential is exactly the one found by Myers
neglecting the back--reaction of the external electric field on
the geometry. This potential has two critical points which are placed
at $\theta=0$ and $r=r_{\pm}$ with
\eqn
r_{\pm}^{\ 2}=\frac{1}{2B^2}
\left(1\pm\sqrt{1-\left(2\pi\a'nB^2\right)^2}\right)\ .
\eeqn
The point $r_-$ is a minimum of the potential while the other
point at $r_+$ is a saddle point.
For small $B$, more precisely for $ B\ll 1/\sqrt{\a' n})$, we have
\begin{equation}
r_{-}\simeq \pi \alpha ^{\prime }nB\ ,
\ \ \ \ \ \ \ 
r_{+}\simeq\frac{1}{B}\ .
\end{equation}
It is simple to check that both extrema are local
minima with respect to variations of $\theta $, and therefore the
full geometry induced by the external electric field confines the
probe to the core of the flux--brane. We recall that, if one
neglects the backreaction and follows the computation of Myers, one
finds a potential with flat directions, corresponding to moving
the probe in the directions orthogonal to the electric field. The
full flux--brane geometry stabilizes the probe, effectively
setting $\theta =0$ ($\Lambda =1$). Therefore $r_{-}$ is a true
minimum of the full potential. This analysis exactly reproduces
the one from supergravity in section 3.2.3, suggesting that the
gravity solution with $r_H=r_+$ is not a minimum of the
gravitational action. The true minimum is unstable through quantum
tunneling, but this process is suppressed in the decoupling limit.
In fact we have already seen that, in the decoupling limit, the
local maximum is not seen anymore, making the local minimum
global. We shall confirm these expectations in the following.

Let us also note that,
as one increases the electric field $ B $, the two extrema $r_{-}$
and $r_{+}$ approach each other, and above a critical electric
field
\begin{equation}
B_{\mathrm{crit}}^{2}=\frac{1}{2\pi \alpha ^{\prime }n}
\end{equation}
the potential does not exhibit any extrema. This analysis
precisely matches the one from supergravity, which was discussed
in section 3.2.1.

Finally, let us note that the above results, although derived in
the flux $7$--brane case, are actually valid for a general flux
$p$--brane, and this can be checked by expanding the flux brane
background fields around its core.

\subsection{The probe potential}

In this section we repeat the above probe computation in the
general background of the dielectric brane. In particular, we will
specialize to the decoupled geometry (\ref{decoupling}), which is
most relevant for the field theory/gravity duality. In this case,
the RR form fields ${\mathcal{A}}_{1}$ and ${\mathcal{A}}_{3}$ are
both non--vanishing, and the forms ${\mathcal{A}}_{5}$ and
${\mathcal{ A}}_{7}$ in the Wess--Zumino part of the probe action
are determined by electro--magnetic duality. More precisely, if
one defines the field strength
${\mathcal{F}}_{p+2}=d{\mathcal{A}}_{p+1}$, and the gauge
invariant field strength
$\widetilde{{\mathcal{F}}}_{p+2}={\mathcal{F}}_{p+2}-H\wedge
{\mathcal{A}}_{p-1}$,  one has\footnote{Throughout the paper we
are using the following convention for Hodge duality. If
$A,B\,$are $p$--forms, then $A\wedge \star B=(A\cdot B)\,\omega $,
where $\omega $ is the volume form and $A\cdot B$ is the inner
product of the forms.} 
\eqn 
\widetilde{\mathcal{F}}_{6} =\star
\widetilde{\mathcal{F}}_{4}\ , \ \ \ \ \ \ \ \ \ \
\widetilde{\mathcal{F}}_{8} =-\star \widetilde{\mathcal{F}}_{2}\ .
\eeqn 
For the background (\ref{decoupling}), a rather tedious
computation gives the
following result for ${\mathcal{A}}_{5}$ and ${\mathcal{A}}_{7}$
\eqn 
{\mathcal{A}}_{5} =\a'^{2}\,a_{5}\,\epsilon \left(
\bM^{5}\right)\  , \ \ \ \ \ \ {\mathcal{A}}_{7} =
\a'^{3}\,a_{7}\,\epsilon \left( \bM^{5}\right) \wedge \epsilon
\left( S_{2}\right)\ , 
\eeqn 
where $a_{5}$ and $a_{7}$ are
functions only of $r,\theta $ and are given by 
\eqn 
\arr{c}
\displaystyle{a_{5} =-\frac{1}{H}-\frac{Ba}{\kappa }\sin ^{2}\theta}\ , 
\spa{0.5}\\
\displaystyle{a_{7} =\frac{2}{3}BU^{3}\cos ^{3}\theta -} \spa{0.5}\\
\displaystyle{-2m_{0}B\cos \theta \left[ -\frac{\sin ^{2}\theta
\left( 1-a\kappa B\sin^{2}\theta \right) }{\Delta }+\left(
1-a\kappa B\right) \left( 1-\frac{1}{3} \cos ^{2}\theta \right)
\right] }\ . 
\earr 
\eeqn 
We recall that the constant $\kappa$, the limit of $\a'\cosh{\a}$ in
the decoupling limit, is given in (\ref{kappa}) and that the magnetic
field parameter $B$ is given in (\ref{Bdecoupl}).
Finally, it is convenient to express the Kalb--Ramond field as
\eqn 
\arr{c}
\displaystyle{{\cal B}=\a'b\,\epsilon \left( S_{2}\right)}\ , 
\spa{0.4}\\
\displaystyle{b =-\frac{2m_{0}\kappa\cos^{3}\th}{\Delta}
\left[B\kappa+\frac{a-a^2B\kappa}{U^{2}}\right]}\ . 
\earr
\eeqn

For static configurations the potential energy $V=V_{BI}+V_{WZ}$
of the brane probe is given by the Born--Infeld and Wess--Zumino
pieces 
\eqn 
\arr{c} 
\displaystyle{V_{BI} =
\frac{V_{4}}{4\pi^{3}g_{YM}^{2}}
\frac{\Sigma ^{1/2}}{H}
\sqrt{\Sigma HU^{4}\cos^{4}\theta +\left( b+\pi n\right) ^{2}}} \ ,
\spa{0.5} \\
\displaystyle{V_{WZ} =\frac{V_{4}}{4\pi ^{3}g_{YM}^{2}}\left[
-a^{\phantom{1}}_{7}+a_{5}\left( b+\pi n\right) \right]}\ . 
\earr
\label{add-1000}
\eeqn

The dimensionless quantities which parameterize the decoupled
geometry solution are $N$ and $g_{YM}^{2}B$. As explained in
section 4, we shall work in the small $B$ regime $ g_{YM}^{2}B\ll 1$,
i.e. within the scope of the gravitational approximation. In this
limit, $U_0$ and $(U_0-a)/U_0$ are given by (\ref{U0,a}), and
the constant $\kappa$ simplifies to
\begin{equation}
\kappa \simeq \frac{1}{\pi B^{2}N}\ .
\end{equation}
The functions $a_{5}$, $a_{7}$ and $b$ are then expressed in terms
of the physical quantities $N$, $B$ and $g_{YM}^{2}$ and simplify
to (here we have only kept the first change in the fields due to
$B$) 
\eqn 
\arr{rcl} 
a_{5} &\simeq &
\displaystyle{-\frac{4\pi}{g_{YM}^{2}N}\, U^{3}+\frac{4\pi
^{3}NB^{2}}{g_{YM}^{2}}\, U\cos ^{2}\theta +\cdots}\ , \spa{0.5}\\
a_{7} &\simeq &
\displaystyle{\frac{2}{3}BU^{3}\cos ^{3}\theta +\cdots}\ , \spa{0.5}\\
b &\simeq &
\displaystyle{-\frac{1}{4\pi }NBg_{YM}^{2}\cos^{3}\theta +\cdots}\ . 
\earr 
\eeqn

\subsubsection{Potential for large $U$}

Now let us analyse the potential $V$ for large values of the
radial coordinate $U$. An expansion in powers of $1/U$ yields
(keeping, for each term in the expansion in $U$, the leading term
in $B$)
\begin{equation}
V=\frac{2V_{4}}{ng_{YM}^{2}}\left[ v^{4}-\frac{2}{3}\left( nB\right) v^{3}+
\frac{1}{4}\left( nB\right) ^{2}\rho ^{2}+\cdots \right] \ ,
\label{add-200}
\end{equation}
where 
\eqn 
v =\frac{U}{2\pi }\cos \theta\ ,
\ \ \ \ \ \ \ \ \
\rho =\frac{U}{2\pi }\sin \theta\ , 
\label{add-600} 
\eeqn
are the radial coordinates along the $\theta =0$ three--plane and the 
$\theta =\pi /2$ two--plane, respectively. Note that the above
potential has the same form as the \textit{improved} Myers
potential (\ref{add-100}). In fact, one can obtain the leading
terms in the potential $V$ in (\ref{add-200}) by starting with the
Myers potential (\ref{add-100}) and taking the decoupling limit
$\alpha ^{\prime }\rightarrow 0$ with $U=r/\alpha ^{\prime}$
fixed. The $\rho ^{2}$ mass term  in $V$ then comes from the
expansion of $\Lambda ^{1/2}$.

The potential $V$ can be
understood in relation to the dual effective low energy action
which governs the physics on the $n$ $D4$--branes  in the presence
of a flux $7$--brane. From the field theory point of view, the
gravity construction in section $3$ corresponds to a twisted
compactification of the $(2,0)$ low energy conformal field theory
on a stack of $M5$--branes. More precisely, recall that the usual
compactification of the $(2,0)$ theory on a circle flows in the IR
to the maximally supersymmetric $U\left( N\right)$ Yang--Mills
theory on the $D4$--branes worldvolume. The gravity analysis of
section $3$ corresponds to a Scherk--Schwarz
circle$\,$compactification \cite{Scherk:1979} which breaks
supersymmetry by twisting the fields with an $R$--symmetry
rotation (breaking the $R$--symmetry group $SO\left( 5\right)$ to
a $SO\left( 3\right) \times SO\left( 2\right) $ subgroup, which is
also evident in the supergravity solution).

To analyze the above reduction in more detail, it is convenient to
first consider the analogous situation of the Scherk--Schwarz
reduction on a circle of ${\mathcal{N}}=4$, $D=4$ SYM. The
situation is similar since both theories are superconformal with
the same number of supercharges and with similar $R$--symmetry
groups, whose action rotates the scalars of the theories. We will
use in the sequel this analogy since the ${\mathcal{N}}=4$, $D=4$
SYM Lagrangian is known, whereas the $\left( 2,0\right) $ theory
Lagrangian is only known for the abelian case \cite{Perry:1997,
Bandos:1997, Aganagic:1997}.

Divide the scalars of the theory in $\Phi ^{m}$, $m=1,2$ and $\Phi
^{A}$, $ A=3,\cdots ,6$. As we go around the compactification
circle, we rotate the scalars $\Phi ^{m}$ in the $1$--$2$ plane by
an angle $2\pi BR_{c}$, while leaving the others fixed (here
$R_{c}$ is the compactification radius). The non trivial part of
the Scherk--Schwarz reduction for the bosonic fields comes from the
kinetic term for the $\Phi ^{m}\,$scalars in the compact direction
\begin{equation}
-\frac{1}{2g_{YM}^{2}}{\mathrm{Tr}}\left( D_{c}\Phi ^{m}D_{c}\Phi
^{m}\right) \rightarrow
-\frac{R_{c}}{g_{YM}^{2}}{\mathrm{Tr}}\left( \frac{1}{2}\,
B^{2}\Phi ^{m}\Phi ^{m}+iB\epsilon _{mn}\Phi ^{m}\left[ \Phi
^{n},\Phi ^{c}\right] \right) ,  \label{add-300}
\end{equation}
where the scalar $\Phi ^{c}$ comes from the dimensional reduction
of the gauge field, and where $g_{YM}^{2}/R_{c}$ is the square of the
Yang--Mills coupling for the $3$--dimensional reduced theory.

Let us move back to the $\left( 2,0\right) $ compactification (now
$A=3,4,5$ since there are only $5$ scalars). At low energies, we
expect the deformed Lagrangian to be
\begin{equation}
{\mathcal{L}}_{SYM}-\frac{1}{g_{YM}^{2}}{\mathrm{Tr}}\left(
\frac{1}{2}\, B^{2}\Phi^{m}\Phi ^{m}+\frac{i}{3}\, B\epsilon
_{ABC}\Phi ^{A}\left[ \Phi ^{B},\Phi ^{C} \right] \right) .
\label{add-400}
\end{equation}
The mass term for the scalars $\Phi ^{m}$ above is exactly the
same as in (\ref{add-300}), if we consider a standard kinetic term
for the scalars of the underlying $\left( 2,0\right) $ theory. The
other term in (\ref{add-400}) is the Myers coupling, and is
similar to the second term in (\ref{add-300}). The difference in
the index structure comes from the fact that the gauge field in
the $\left( 2,0\right) $ theory is not a $1$--form but rather a
matrix--valued $2$--form, whose coupling to the matter fields is
not known in the non--abelian case. Therefore we predict that the
Scherk--Schwarz reduction of the non--abelian $\left( 2,0\right) $
theory gives rise to the Myers coupling in (\ref{add-400}).

To conclude this discussion we match exactly the Lagrangian
(\ref{add-400}) to the probe potential (\ref{add-200}),
following \cite{PolcStra:00}. This is easily done by the ansatz
\eqn 
\Phi ^{1}+i\Phi ^{2} =e^{i\phi }\, \rho \cdot
{\mathbf{1}}_{n\times n}\ ,  \ \ \ \ \ \ \ \  \Phi ^{A}
=\frac{2}{n}\, v\cdot \alpha _{n\times n}^{A}\ ,
\label{add-500}
\eeqn 
where $\alpha _{n\times n}^{A}$ is the $n$--dimensional
representation of the $SU\left(2\right)$ algebra, normalized to
$4\alpha ^{A}\alpha ^{A}=\left( n^{2}-1\right) \cdot $
$\mathbf{1}$ (or $\left[ \alpha ^{A},\alpha ^{B} \right]
=i\epsilon _{ABC}\alpha ^{C}$). The coefficients in
(\ref{add-500}) are fixed by the conditions
\begin{equation}
\Phi ^{m}\Phi ^{m}=\rho ^{2}\cdot {\mathbf{1}}_{n\times n}\ ,
\ \ \ \ \ \ \ \ \ \ \ 
\Phi ^{A}\Phi ^{A}=v^{2}\cdot
{\mathbf{1}}_{n\times n}\ .
\end{equation}

\subsubsection{Potential inside the dielectric brane}

We proceed to analyse the probe potential around the center of the
dielectric brane. More precisely, we will expand the potential
around $ U=U_{0}$ and $\theta =\pi /2$. Recall that inside the
dielectric brane the radial coordinate in the three--plane is
$U_{0}\cos \theta $. On the other hand, the radial coordinate in
the transverse two--plane needs to be modified. It turns out, as
one can guess from the coordinate transformation
(\ref{spheroidal}), that the appropriate definitions for the
radial coordinates $v$ and $ \rho $ are now 
\eqn 
v =\frac{U}{2\pi}\cos \theta \ , 
\ \ \ \ \ \ \ 
\rho  =\frac{\sqrt{U^{2}-U_{0}^{\ 2}}}{2\pi}\, \sin \theta\ . 
\eeqn
In the limit of large $U$ considered before we recover
(\ref{add-600}). Moreover, in terms of the above radial variables, the
probe potential inside the dielectric brane has \textit{exactly} the
same form, to leading order in the deformation parameter $B$, as in
(\ref{add-200}).

Now we can easily see, from the form of the potential $V$ around the
center of the dielectric sphere, that the probe has one stable minimum
at  
\eqn
U=U_{0}\ ,
\ \ \ \ \  
\cos{\th}=\frac{\pi nB}{U_0}\simeq\frac{n}{N}\ .
\eeqn
This correspond to a dielectric shell of $n$ $D4$--branes inside
the larger shell of $N$ $D4$--branes associated to the background
geometry. This configuration corresponds to a vacuum of
(\ref{add-400}) formed by a reducible $SU\left(2\right) $
representation. Moreover, as the probe approaches the outer shell
at $U=U_{0}$, $\theta=0$, one finds a second minimum corresponding
to the fusion of the probe to the background dielectric brane, as
can be seen by the plot of the full potential (\ref{add-1000}) in
Figure 4. Very close to the dielectric brane there will be
curvature corrections to the geometry, which will correct the
potential. However we do not expect this qualitative behavior to
change.

\begin{figure}[t]
\begin{picture}(200,150)(0,0)
{\centerline{\psfig{figure=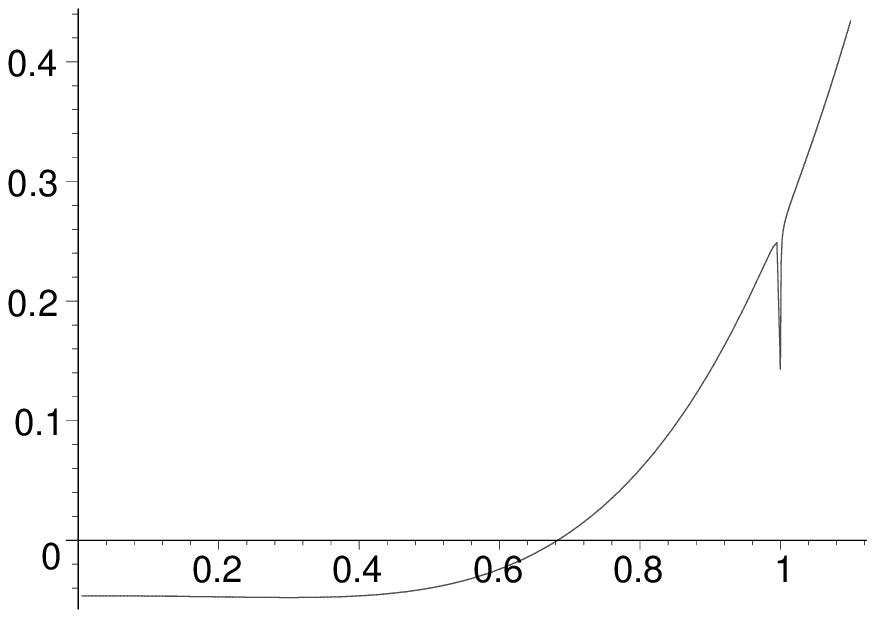,width=7cm}}}
\put(-125,14){$x$} 
\put(-145,-10){\vector(1,-3){10}}
\put(-275,-10){\vector(-1,-3){10}}
\end{picture}
\begin{picture}(300,150)(0,0)
\put(50,0){\psfig{figure=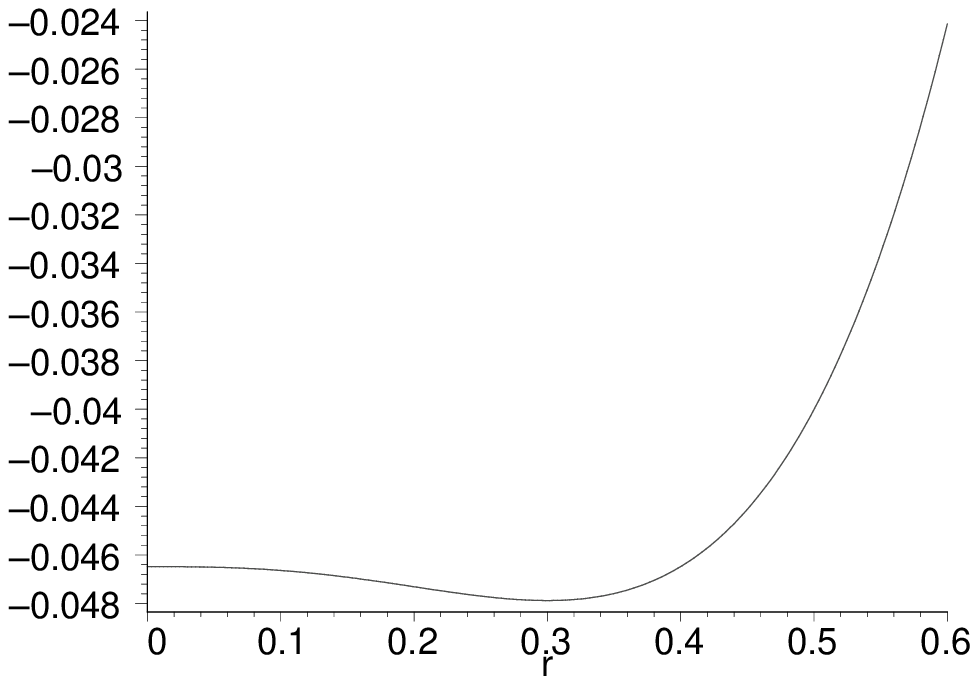,width=5cm}}
\put(270,0){\psfig{figure=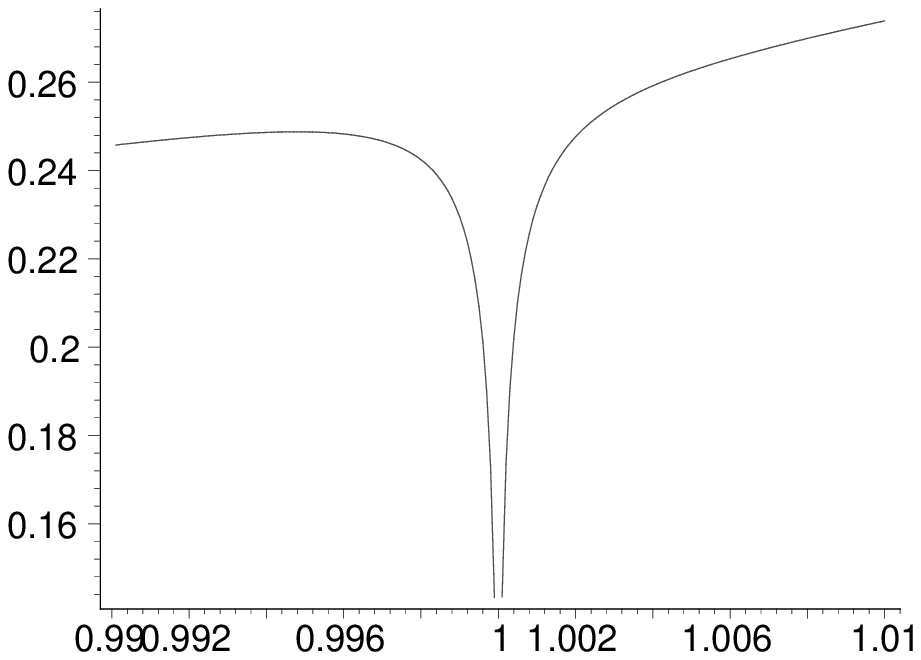,width=5cm}}
\put(127,1){$_x$}
\put(348,-2){$_x$}
\put(415,1){\phantom{\rule{6mm}{2mm}}}
\end{picture}
\caption{\small{The two minima of the potential for $g^2_{YM}B=0.01$ 
and $n/N=0.3$. In the top graph the dimensionless radial coordinate 
$x$ is defined, for $0\le x<1$, by $U=U_0$, $x=\cos{\th}$ and it is 
the radial coordinate inside the dielectric sphere. For $x>1$, we have 
$x=U/U_0$ and $\th=0$ as appropriate for the radial coordinate outside
the dielectric sphere on the $\th=0$ three--plane. The other two plots 
zoom into the minima whose interpretation is given in the main text.}}
\end{figure}

To conclude we summarize our findings of this section. Motivated
by the gravity construction, the dual field theory is defined as a
Scherk--Schwarz compactification of the $\left( 2,0\right) $
theory. From the point of view of the effective Lagrangian, the
Scherk--Schwarz reduction introduces the IR relevant couplings in
(\ref{add-400}) defined at the compactification energy scale. The
probe computation, based on the gravity solution
(\ref{decoupling}), matches exactly the Lagrangian (\ref{add-400})
in the region of small curvature, which corresponds to the
strongly coupled regime of the undeformed SYM theory. This region
actually extends to the interior of the background dielectric
brane, where again the probe potential has exactly the same form
to leading order in $B$. As one approaches the dielectric brane,
the supergravity approximation breaks down. This corresponds to
the deep infrared regime of the theory, where the deformation terms in
(\ref{add-400}) become relevant. The corrections to the classical
VEV of the scalars can be seen from the gravity prediction for the
dielectric radius $U_{0}$ which differs from its classical value
$\pi NB$ by $g_{YM}^{2}B$ corrections.

\section{Conclusion}

In this paper we have shown the existence of a new type of branes
is String and M-theory. These are flux $p$--branes with
$(p+1)$--dimensional Poincar\'e invariance. They are characterized
by a flux of an associated form field strength through the
transverse space to the brane. To find the analytic solution for
the flux branes would require solving a system of differential
equations that can be casted in the form of interacting Liouville
systems. The asymptotics  of the flux brane geometry are the same
as the asymptotics of a particular singular analytic solution.

The flux branes are non-supersymmetric and non-stable vacua of
String and M-theory. We argued that for a RR flux $(p+3)$--brane,
this vacuum can be stabilized by placing $N$ D$p$--branes on the
flux brane. This leads to the expansion of the D--brane to a
dielectric brane, rendering the system classically stable but unstable 
under quantum tunneling. After taking the decoupling limit the
configuration is stable both classically and quantum mechanically.

With the help of the M-theory Ka\l u\.{z}a--Klein reduction to the
type IIA theory, we were able to construct the gravity solution
for a D4--brane expanded into a dielectric 2--sphere due to the
presence of a flux 7--brane. This is the first exact gravity
solution for the dielectric effect in the literature. Previous
work has considered a spherical configuration of D3--brane charge,
with a uniform source of 5-brane charge, and then solved the
gravity equations perturbatively far away from the sources
\cite{PolcStra:00}. A new venue of research is to consider $N$
D$3$-branes placed on RR and/or NSNS flux 6--branes. This
configuration should be of the type studied in \cite{PolcStra:00}.

The analysis of the probe potential in the decoupled geometry gave us
some hints for what the Scherk--Schwarz reduction of the M5--branes
$(2,0)$ low energy conformal field theory should be. We were
successful in reproducing the mass term in the dual field theory that
arises from the reduction, but the precise origin of the Myers cubic term
remains unclear. We think this issue deserves further investigation
since the full `non--abelian' action for the M5--branes effective
theory remains unknown. 

\section*{Note Added}
While this work was in progress two pre--prints
\cite{Saffin,GutpStro} appeared that have some overlap with the
material presented in section 2 of this paper. In particular,
they investigate the asymptotics \cite{Saffin,GutpStro} and the
`near--core' \cite{GutpStro} behavior of the solutions to the flux
brane equations (\ref{fgeqns}) that we had independently derived.

\section*{Acknowledgments}
We are grateful to Costas Bachas, Dominic Brecher, Gary Gibbons, 
Michael Gutperle, Elias Kiritsis, Malcolm Perry and Harvey Reall for
helpful discussions. The work of M.S.C. was supported by a Marie Curie
Fellowship under the European Commission's Improving Human
Potential programme. The work of C.H. was supported by FCT
(Portugal) through grant no. PRAXIS XXI/BD/13384/97. L.C. was
funded by an European Postdoctoral Institute fellowship.


\begin{thebibliography}{99}

\bibitem{HoroStro}
G.T. Horowitz and A. Strominger, {\em Black Strings and p-branes},
Nucl. Phys. {\bf B360} (1991) 197.

\bibitem{Polc:95}
J. Polchinski, {\em Dirichlet-Branes and Ramond-Ramond Charges},
Phys. Rev. Lett. {\bf 75} (1995) 4724 [hep-th/9510017].

\bibitem{StroVafa}
A. Strominger and C. Vafa, {\em Microscopic Origin of the
Bekenstein-Hawking Entropy}, Phys. Lett. {\bf B379} (1996) 99
[hep-th/9601029].

\bibitem{Mald:97}
J.~M.~Maldacena, {\em The Large N Limit of Superconformal Field
Theories and Supergravity}, Adv. Theor. Math. Phys. {\bf 2} (1998)
231 [hep-th/9711200].

\bibitem{GKP}S.S. Gubser, I.R. Klebanov and A.M. Polyakov,
{\em Gauge Theory Correlators from Non-Critical String Theory},
Phys. Lett. {\bf B428} (1998) 105 [hep-th/9802109].

\bibitem{Witten}E. Witten,
{\em Anti De Sitter Space And Holography}, Adv. Theor. Math. Phys.
{\bf 2} (1998) 253 [hep-th/9802150].

\bibitem{Melv:64}
M. Melvin, {\em Pure Magnetic and Electric Geons}, Phys. Lett.
{\bf 8} (1964) 65.

\bibitem{Melv:65}
M. Melvin, {\em Dynamics of Cylindrical Electromagnetic
Universes}, Phys. Rev. {\bf 139} (1965) B225.

\bibitem{Ernst}
F. Ernst, {\em Black Holes in a magnetic universe}, J. Math. Phys.
{\bf 17} (1976) 54.

\bibitem{GarfMelv}
D. Garfinkle and M. Melvin, {\em Generalized magnetic universe
solutions}, Phys. Rev. {\bf D50} (1994) 3859.

\bibitem{Gibbons:2001}
G.~W.~Gibbons and C.~A.~R.~Herdeiro, {\em The Melvin universe in
Born-Infeld theory and other theories of  non-linear
electrodynamics}, [hep-th/0101229].

\bibitem{Dowker:93}
F.~Dowker, J.~P.~Gauntlett, D.~A.~Kastor and J.~Traschen,
{\em Pair Creation of Dilaton Black Holes}, Phys.\ Rev.\ {\bf D49} 
(1994) 2909 [hep-th/9309075].

\bibitem{Dowker:95}
F.~Dowker, J.~P.~Gauntlett, G.~W.~Gibbons and G.~T.~Horowitz, 
{\em The Decay of magnetic fields in Kaluza-Klein theory}, Phys.\ Rev.\
{\bf D52} (1995) 6929 [hep-th/9507143].

\bibitem{Dowker:96}
F.~Dowker, J.~P.~Gauntlett, G.~W.~Gibbons and G.~T.~Horowitz, 
{\em Nucleation of $P$-Branes and Fundamental Strings}, Phys.\ Rev.\
{\bf D53} (1996) 7115 [hep-th/9512154].

\bibitem{RussoTsey}
J.~G.~Russo and A.~A.~Tseytlin, {\em Green-Schwarz superstring
action in a curved magnetic Ramond-Ramond background}, JHEP {\bf
9804}, 014 (1998) [hep-th/9804076].

\bibitem{CostaGutp:01}
M.~S.~Costa and M.~Gutperle, {\em The Kaluza-Klein Melvin Solution
in M-theory}, [hep-th/0012072].

\bibitem{RussoTsey2}
J.~G.~Russo and A.~A.~Tseytlin, {\em Constant magnetic field in
closed string theory: An Exactly solvable model}, Nucl.\ Phys.\
{\bf B448} (1995) 293 [hep-th/9411099].

\bibitem{RussoTsey3}
A.~A.~Tseytlin, {\em Closed superstrings in magnetic flux tube
background}, Nucl.\ Phys.\ Proc.\ Suppl.\  {\bf 49} (1996) 338
[hep-th/9510041].

\bibitem{RussoTsey4}
J.~G.~Russo and A.~A.~Tseytlin, {\em Magnetic flux tube models in
superstring theory}, Nucl.\ Phys.\  {\bf B461} (1996) 131
[hep-th/9508068].

\bibitem{RussoTsey5}
J.~G.~Russo and A.~A.~Tseytlin, {\em Magnetic backgrounds and
tachyonic instabilities in closed superstring theory and
M-theory}, [hep-th/0104238].

\bibitem{JansMukh}
B.~Janssen and S.~Mukherji, {\em Kaluza-Klein dipoles,
brane/anti-brane pairs and instabilities}, [hep-th/9905153].

\bibitem{Myers:99}
R.~C.~Myers, {\em Dielectric-branes}, JHEP {\bf 12} (1999) 022
[hep-th/9910053].

\bibitem{PolcStra:00}
J. Polchinski and  M. Strassler, {\em The String Dual of a
Confining Four-Dimensional Gauge Theory}, [hep-th/0003136].

\bibitem{PilchWarn}
K. Pilch and N. Warner, {\em $N=1$ Supersymmetric Renormalization
Group Flows from IIB Supergravity}, [hep-th/0006066].

\bibitem{FreeMina}
D.Z. Freedman and J.A. Minahan, {\em Finite Temperature Effects in
the Supergravity Dual of the $N=1^*$ Gauge Theory}, JHEP {\bf 01}
(2001) 036 [hep-th/0007250].

\bibitem{PilchWarn2}
K. Pilch and N. Warner, {\em A Class of $N=1$ Supersymmetric RG
Flows from Five-dimensional $N=8$ Supergravity}, Phys. Lett. {\bf
B495} (2000) 215 [hep-th/0009159].

\bibitem{Robi}
M. Taylor-Robinson, {\em Anomalies, counterterms and the ${\cal N}
=0$ Polchinski-Strassler solutions}, [hep-th/0103162].

\bibitem{Town:95}
P.~K.~Townsend, {\em The eleven-dimensional supermembrane
revisited}, Phys. Lett. {\bf B350} (1995) 184 [hep-th/9501068].

\bibitem{Scherk:1979}
J.~Scherk and J.~H.~Schwarz, {\em Spontaneous Breaking Of
Supersymmetry Through Dimensional Reduction}, Phys.\ Lett.\ B {\bf
82} (1979) 60.

\bibitem{Callan}
C.G. Callan, S.S. Gubser, I.R. Klebanov, A.A. Tseytlin, {\em
Absorption of Fixed scalars and the D-brane Approach to Black
Holes}, Nucl. Phys. {\bf B489} (1997) 65 [hep-th/9610172].

\bibitem{GibbMaeda}
G.W. Gibbons and K. Maeda, {\em Black Holes and Membranes in
Higher Dimensional Theories with Dilaton Fields}, Nucl. Phys. {\bf
B298} (1988) 741.

\bibitem{Saffin}
P.M. Saffin, {\em Gravitating Fluxbranes}, [gr-qc/0104014].

\bibitem{GutpStro}
M.~Gutperle and A. Strominger, {\em Fluxbranes in String Theory},
[hep-th/0104136].

\bibitem{PerryMyers}
R.~C.~Myers and M.~J.~Perry, {\em Black Holes In Higher
Dimensional Space-Times}, Annals Phys.\  {\bf 172} (1986) 304.

\bibitem{Brec:96}
J.~C.~Breckenridge, G.~Michaud and  R.~C.~Myers, {\em More D-brane
bound states}, Phys. Rev. {\bf D55} (1997) 6438 [hep-th/9611174].

\bibitem{CostaPapa:96}
M.~S.~Costa and G.~Papadopoulos, {\em Superstring dualities and
p-brane bound states}, Nucl. Phys. {\bf B510} (1998) 217
[hep-th/9612204].

\bibitem{Cvet:96}
M.~Cveti\v c and D.~Youm, {\em Rotating Intersecting M-Branes},
Nucl. Phys. {\bf B499} (1997) 253 [hep-th/9612229].

\bibitem{Sfet:99}
C.~Cs\'aki, J.~Russo, K.~Sfetsos and J.~Terning, {\em Supergravity
Models for 3+1 Dimensional QCD}, Phys. Rev. {\bf D60} (1999)
044001 [hep-th/9902067].

\bibitem{MaldStro}
J.~Maldacena and A.~Strominger, {\em AdS(3) black holes and a
stringy exclusion principle}, JHEP {\bf 12} (1998) 005
[hep-th/9804085].

\bibitem{BrecSaff}
D.~Brecher and P.~M.~Saffin, 
{\em A note on the Supergravity Description of Dielectric Branes},
[hep-th/0106206].

\bibitem{Hawking:95}
S.~W.~Hawking and G.~T.~Horowitz, {\em The Gravitational
Hamiltonian, action, entropy and surface terms}, Class.\ Quant.\
Grav.\ {\bf 13} (1996) 1487 [gr-qc/9501014].

\bibitem{Itzh:98}
N.~Itzhaki, J.~M.~Maldacena, J.~Sonnenschein, S.~Yankielowicz,
{\em Supergravity and The Large N Limit of Theories With Sixteen
Supercharges}, Phys. \ Rev. \ {\bf D58} (1998) 046004
[hep-th/9802042].

\bibitem{Perry:1997}
M.~Perry and J.~H.~Schwarz, {\em Interacting chiral gauge fields
in six dimensions and Born-Infeld  theory}, Nucl.\ Phys.\ B {\bf
489} (1997) 47 [hep-th/9611065].

\bibitem{Bandos:1997}
I.~Bandos, K.~Lechner, A.~Nurmagambetov, P.~Pasti, D.~Sorokin and
M.~Tonin, {\em Covariant action for the super-five-brane of
M-theory}, Phys.\ Rev.\ Lett.\  {\bf 78} (1997) 4332
[hep-th/9701149].

\bibitem{Aganagic:1997}
M.~Aganagic, J.~Park, C.~Popescu and J.~H.~Schwarz, {\em
World-volume action of the M-theory five-brane}, Nucl.\ Phys.\ B
{\bf 496} (1997) 191 [hep-th/9701166].
\end{thebibliography}
\end{document}